\documentclass[aps]{revtex4}

      \newcommand{\beq}{\begin{equation}} 
      \newcommand{\eeq}{\end{equation}} 
      \newcommand{\beqa}{\begin{eqnarray}} 
      \newcommand{\eeqa}{\end{eqnarray}} 
      \newcommand{\ket}[2]{|#1\rangle_{#2}} 
      \newcommand{\bra}[2]{{}_{#2}\langle #1|} 
      \newcommand{\kett}[3]{|#1\rangle_{#2}^{#3}} 
      \newcommand{\braa}[3]{{}_{#2}^{#3}\langle #1|}

      \def\<{\langle} 
      \def\>{\rangle}

      \def\opone{\leavevmode\hbox{\small1\kern-3.8pt\normalsize1}} 
       
      \newcommand{\complex}{{\kern .1em {\raise .47ex\hbox {$\scriptscriptstyle 
      |$}}\kern -.4em {\rm C}}} 
      \newcommand{\real}{{{\rm I} \kern -.19em {\rm R}}} 

\begin{document} 
  \title{Error filtration and Entanglement Purification 
for quantum communication}

\author{N. Gisin}
\affiliation{Group of Applied Physics, 20, rue de l'Ecole-de-M\'edecine, 
      CH-1211 Geneva 4, Switzerland}
\author{N. Linden}
\affiliation{Dept. of Mathematics, University of Bristol, University Walk, 
      Bristol BS8 1TW, U.K.}
\author {Serge Massar}
\affiliation{Physique Th\'{e}orique, {C.P.} 225, Universit\'{e} Libre 
de Bruxelles, Boulevard du Triomphe, 1050 Bruxelles, Belgium}
\affiliation{Centre for Quantum Information and Communication, {C.P.}
  165/59, Universit\'{e} Libre de Bruxelles, Avenue F. D. Roosevelt  
  50, 1050 Bruxelles, Belgium}  
\author{S. Popescu}
\affiliation{H. H. Wills Physics Laboratory, University of Bristol, Tyndall 
      Avenue, 
      Bristol BS8 1TL, U.K.}
\affiliation{Hewlett-Packard Laboratories, Stoke Gifford, Bristol BS12
  6QZ, U.K.} 
\begin{abstract}
The key realisation which lead to the emergence of the new 
      field of quantum information processing is that quantum mechanics, 
      the theory that describes microscopic particles, allows the 
      processing of information in fundamentally new ways. But just as 
      in classical information processing, errors occur in quantum 
      information processing, and these have to be corrected. A 
      fundamental breakthrough was the realisation that quantum error 
      correction is in fact possible. However most work so far has not 
      been concerned with technological feasibility, but rather with 
      proving that quantum error correction is possible in principle. 
      Here we describe a method for {\it filtering} out errors and 
      {\it entanglement purification} which is 
      particularly suitable for quantum communication. Our method is 
      conceptually new, and, crucially, it is easy to implement in a 
      wide variety of physical systems with present day technology and 
      should therefore be of wide applicability.
\end{abstract}

\maketitle

\section{Introduction} 

When quantum communication was first 
      proposed it was felt that interactions of the system with the 
      environment, and the consequent loss of information into the 
      environment, would produce errors which would be un-correctable 
      even in principle. However it was discovered using two independent 
      approaches, namely error correction codes for quantum memories 
      \cite{Shor,Steane} and entanglement purification 
      \cite{errorcorrection}, that quantum error correction is in fact 
      possible. This transformed the field from an intellectual game 
      into a potentially revolutionary new technology. 

      In these pioneering papers, and their extensions 
      (see for instance \cite{B96,K96,Z97,L98,symm}), 
      the authors were not concerned with immediate 
      technological feasibility, but rather with proving a point of 
      principle. And the difficulty with all these protocols is that 
      in order to be implemented they require controlled interactions 
      between many particles. This is technically impractical at present 
      and it is likely to remain so for the foreseeable future. For this 
      reason several authors have proposed methods for error correction 
      which are comparatively simpler to implement 
      \cite{Braunstein,LloydSlotine,Bouwmeester,Zeilinger,Duan}, one of 
      which has recently been demonstrated \cite{Zeilinger2}. However all 
      these methods require special resources such as squeezed 
      states \cite{Braunstein,LloydSlotine}, or particular entangled 
      states \cite{Bouwmeester,Zeilinger} or optical memories \cite{Duan} 
      that are at the limit of present technology even for proof of 
      principle, let alone for practical schemes. 

      Here we describe a different approach to dealing with errors that 
      works in a conceptually different way to existing error correction 
      methods. Our method realizes {\it error filtration} and is 
      particularly useful for quantum communication. It can easily be 
      implemented with present day technologies and have 
      therefore the potential to make a significant impact on the 
      nascent field of quantum communication. In the longer term, since our 
      methods are applicable to any type of communication, they may find 
      uses inside quantum computers and other quantum devices, if and 
      when these are built.

      The difference between error correction and error filtration is 
      the following. In error {\em correction} the aim is actively to 
      correct the errors that occur during transmission so that the 
      decoded signal is as close as possible to the emitted signal. In 
      error {\em filtration} the aim is to detect with high probability 
      when an error has occurred, and in that case to discard the 
      signal. In effect, what this method does is transform 
      a general error (phase noise, depolarisation, etc...) into an 
      erasure, which is far more benign.

      We present two methods: one for improving a channel which is 
      useful for distributing arbitrary states (channel multiplexing), 
      the other for distributing a standard entangled state (source 
      multiplexing). Of course, the first method can also be used for 
      distributing entangled states, but when we {\em know} what 
      entangled state we want we can also manipulate the source. Thus our 
      methods also provide 
      new ways of {\it purifying} entangled states. 
Several 
      extensions of these schemes are presented in the 
      Appendix.

      The simplicity of implementation of error filtration is underlined 
      by an experiment, reported elsewhere \cite{QCrypto}, in which a 
      particular scheme is implemented.

      \section{Basic principle} 

      The central idea of the first method is what we will call {\em 
      channel multiplexing}, ie. to use more transmission channels than 
      the minimum necessary to send the quantum state. 

      The main concept of our method is extremely simple and we illustrate it by 
      means of 
      an example. Consider a 
      quantum system that propagates through a communication channel, 
      such as a photon going through an optical fiber. During the 
      propagation errors can occur; it is these errors that we want to 
      identify and get rid of. Our method consists of replacing the 
      original single communication channel by an interferometer 
      consisting of two communication channels in parallel. Now, instead 
      of sending the photon through the original single channel, we send 
      it in a superposition of two states, one going through each arm of 
      the interferometer, i.e. in a superposition of going through the 
      two channels (see Fig 1). 

      The errors in each channel are arbitrary, i.e. we impose 
      no restrictions on the properties of the individual 
      channels. We will however arrange things so that the errors 
      on the different channels are independent. This is a simple 
      matter of engineering. (For instance we can always 
      increase the space separation of the channels) 

      When emerging from the two channels, the two wave-packets 
      interfere. The output beam-splitter is tuned in such a way that 
      when no error occurs, the photon emerges with certainty in one of 
      the output channels; we call this the ``useful" output. That is, 
      we arrange the interferometer so that, in absence of errors, there 
      is complete constructive interference in one output channel and 
      complete destructive interference in the other output channel. 

      In presence of errors, the output in the useful channel is better 
      than if we had not multiplexed. The reason for this may be 
      understood as follows: 

      Let us first consider that all the channels are similar, i.e. that 
      they produce the same amount of errors. The total probability for 
      an error is the same whether we send the photon only through one 
      single channel, as in the original scheme, or in a superposition 
      of going through the two channels. (Indeed, we do not consider two 
      photons, one going through one channel, the other going through 
      the other channel, each accumulating errors, but a single photon, 
      going either in one channel or in the other.) 

      There are two equivalent but complementary ways of describing 
      noise \cite{NC,YA}. The first is to view the system as undergoing 
      an evolution which depends on random parameters. The second is to 
      describe the system and its environment as a single combined 
      system. This combined systems evolves unitarily, and the noise 
      manifests itself as entanglement between the system and the 
      environment. In the second description, which we adopt here, 
      whenever an error occurs 
      there is a registration in the environment. 
      Due to the fact that 
      the channels are independent, (as described above), when an error 
      affects the photon going through the first channel, it is the 
      state of the first channel that it is affected (i.e. that 
      registers the fact that the error occurred). Similarly, when the 
      photon goes through the second channel and an error occurs, it is 
      the state of the second channel that is affected.

      Thus, whenever an error occurs, one could, in 
      principle, by looking at the state of the two channels, find out 
      which channel the photon went through. This causes the original 
      superposition (of the photon going through both channels) to 
      collapse. Hence, at the output beam-splitter, the two wave-packets 
      no longer interfere, and the photon may sometimes end up at the 
      useful output, while other times it can end up at the other output. 
      In this latter case we {\it know} that an error occurred - 
      otherwise the photon couldn't have ended up there - so we discard 
      it. We thus get rid of some errors. And since multiplexing didn't 
      increase the total number of errors, we end up with fewer errors.

      Once this simple principle is understood, it is clear that many 
      variations are possible. 
      For example, in Fig. 1 we view the channels as being different 
      optical fibers. The 
      same scheme works for free space propagation of photons, or for 
      propagation of electrons through mesoscopic wires. Also two 
      ``separate channels" need not be implemented as two different 
      physical objects. For example, one can use a single optical 
      fiber and represent the different channels 
      by different ``time bins", see Fig 2 and \cite{2photonsource99}. In this 
      case, of 
      course, one has to take care to 
      make the channels independent; making the time bins sufficiently separated 
      in 
      time can easily achieve this goal. 

      Obviously, we can use a greater degree of multiplexing, i.e., 
      replace one channel not by two but by many. This further decreases the 
      amount of noise in the useful output. We also can cut a 
      channel into pieces and use filtration on each piece separately, 
      in series, one after the other. 
      It is also not critical that the different channels have the same 
      amount of noise (as we assumed for simplicity above).

      It is also clear that there is a lot of freedom in the relative 
      phase of the superposition of the photon wave-packets going 
      through the different channels. All that is required is that, in 
      the absence of noise, at the output there is constructive 
      interference so that the photon emerges with certainty in one 
      particular output channel (the ``useful" output port). This 
      depends however on both the input into the interferometer and on 
      the output. We can arrange any input, and then tune the output 
      accordingly. (In some examples below, we arrange the superposition 
      by Hadamard transform; in other examples we use the Fourier 
      transform; but these are just examples from a large family). 

      Finally it is clear that the method will work independently of how 
      the quantum information is encoded in the particle. For instance 
      it may be encoded in the relative phase between time bins (as in 
      Fig. 2), or an internal degree of freedom such as the polarisation 
      of the photon or spin of an electron. In all cases the method will 
      improve the output signal.

      We now give a number of examples in detail.

      \section{Error filtration for a single particle}

  Let us consider a source which produces a signal encoded 
      in the quantum state of a single particle: $|\psi\rangle = 
      \sum_{l=1}^{S_{tot}} c_l |l\rangle_S$ where $c_l$ is the complex 
      amplitude that the particle is in source channel $l$ and 
      $|l\rangle_S$ denotes the state of the particle if it is in source 
      channel $l$. Let us suppose that there are a certain number 
      $T_{tot}$ of transmission channels with $T_{tot} = T S_{tot}$ a 
      multiple of $S_{tot}$. Denote by $|j\rangle_T$ ($j=1,...,T_{tot}$) 
      the state of the particle if it is in transmission channel $j$. 

      During transmission the states $|j\rangle_T$ will be affected by 
      noise. For the sake of illustration we consider the simple, but 
      experimentally relevant (see Fig. 1 and \cite{QCrypto}) case 
      where the noise affecting the different states is 
      phase noise. Furthermore, by construction, we can arrange that the 
      noise in the different channels is independent. As mentioned in 
      the preceding section, this doesn't depend on the intrinsic nature of 
      the channels, but on the way we put the channels together, and it 
      is, in general, rather simple to do. 

      Then the states $\ket jT $ evolve according to $\ket jT \mapsto 
      e^{i\phi_j}\ket jT$ where $\phi_j$ are independent random 
      variables. An equivalent description of the noise is to suppose 
      that during transmission the particle interacts with the 
      environment as follows: 
      \beqa \ket jT \ket 0E \mapsto \ket jT 
      \left( \alpha_j \ket 0E + \beta_j \ket jE\right) \quad , \quad | 
      \alpha_j|^2 + |\beta_j|^2 =1 \label{phase-noise}\eeqa where the 
      environment states $\ket jE$ are orthogonal for different $j$. 
      This represents the physical situation in which the environment 
      starts in the initial state $\ket 0E$ and the interaction causes 
      the state of the environment to be disturbed. The amplitudes 
      $\alpha_j$ and $\beta_j$ describe the amount of disturbance; in 
      general these parameters will depend on $j$. For simplicity we 
      first consider the case that they are the same on all channels, i.e. 
      $\alpha_j$ and $\beta_j$ are independent of $j$. 

      We suppose that each source channel is encoded and decoded 
      separately. We can therefore focus on a particular one, $\ket 1S$, 
      say. The source state $\ket 1S$ is encoded into $T$ transmission 
      channels: \beqa \ket 1S \mapsto U_e \ket1S = {1\over\sqrt{T}} 
      \sum_{j=1}^{T}\ket jT\label{Fourier} \eeqa where we have taken 
      $U_e$ to be the discrete Fourier transform. The noise now occurs, 
      causing the state of system plus environment to change to 
      $ 
      {1\over\sqrt{T}} \sum_{j=1}^{T}\ket jT \left( \alpha \ket 0E + 
      \beta \ket jE\right). 
      $ 
      We decode by performing the inverse Fourier transform: 
      $ 
      \ket jT \mapsto U_d \ket jT = {1\over\sqrt{T}} 
      \sum_{k=1}^{T}e^{2i\pi j(k-1)/T}\ket kR\label{InverseFourier} $, 
      where $\ket kR$ denotes the state in receiver channel $k$. 
      Such encoding and decoding transformations can easily be easily realised 
      using linear elements such as beam splitters and phase 
      shifters\cite{linear}. One relevant figure of merit is 
      simply the number of such elements which are required to realise 
      the encoding and decoding operations. The Fourier transform 
      which we use here for illustrative purposes is not necessarily optimal 
      in this respect. 

      One 
      easily computes that the state after decoding is 
      \beqa \ket 1R 
      \left( \alpha \ket 0E + {\beta \over \sqrt{T}} \ket {\tilde 1} 
      E\right)+{\beta\over {T}} \sum_{j=1}^{T}\sum_{k=2}^{T}e^{2i\pi 
      j(k-1)/T}\ket kR \ket jE,\label{output-state} \label{three3} 
      \eeqa 
      where $\ket {\tilde 1}E ={1\over \sqrt{T}}\sum_{j=1}^{T }\ket jE $ 
      is a normalised state of the environment. We see that this state 
      has a component in the receiver channel $\ket 1R$, which we regard 
      as the ``useful'' signal, and components in all the other receiver 
      channels, $\ket kR$, $k\neq 1$ which we discard. The useful signal 
      is 
      $ 
      \ket 1R \left( \alpha \ket 0E + {\beta \over \sqrt{T}} \ket 
      {\tilde 1} E\right)$. The norm of this state gives the probability 
      that the state appears in the useful channel: 
      $P_{success}=|\alpha|^2 + {|\beta|^2\over T}$. 

      In summary the signal does not always reach the useful receiver 
      channel. But when it does the noise amplitude is reduced by a 
      factor of $\sqrt{T}$. As an illustration consider the case where 
      the number of source states $S_{tot}=2$ and the state to be 
      transmitted is $|\psi \rangle=\ket 1S + e^{i\Phi} \ket {2}S $. The 
      ability to preserve the phase $\Phi$ is a measure of the quality 
      of the transmission. This is often quantified by the visibility 
      $V$ of the interference fringes seen by the receiver if he 
      measures in the $\ket 1R \pm\ket 2R$ basis. One finds $V= T 
      |\alpha|^2 / ( T|\alpha|^2 + |\beta|^2)$ which equals $|\alpha|^2$ 
      for $T=1$ (no multiplexing) and increases monotonically to 1 as 
      the amount of multiplexing $T$ tends to infinity. 

      Above we considered for simplicity that all the channels were 
      identical but the method works equally well if the channels are 
      different. This is an essential property since it shows that the 
      method is robust against perturbations. To prove this suppose that 
      each channel $j$ is characterised by parameters $\alpha_j$ and 
      $\beta_j$ which obey $|\alpha_j|^2 + |\beta_j|^2=1$ and use the 
      same encoding and decoding procedure as above. The amplitude of 
      the quantum state if it ends up in the useful receiver channel 
      (the analogue of the first term of eq. (\ref{three3})) is 
      $$ 
      |1\rangle_R \left( 
      \overline{\alpha} |0\rangle_E + 
      {1\over T} \sum_{j=1}^T \beta_j |j\rangle_E\right) 
      $$ 
      where $\overline \alpha = {1\over T} \sum_{j=1}^T \alpha_j$. The 
      probability that the signal ends up in the useful receiver channel 
      and is unaffected by error is $P_{success\& no\ 
      error}=|\overline \alpha|^2$ whereas the probability that the signal 
      ends up in the useful channel and is affected by error is 
      $P_{success\& error}={1\over T^2} \sum_{j=1}^T |\beta_j|^2 \leq 
      {1\over T} (1 - |\overline \alpha|^2)$. Let us suppose that the 
      $\alpha_j$ all have approximately the same phase (this can easily 
      be arranged by putting a phase shifter in each channel) and that 
      $\overline \alpha$ does not tend to zero as the degree of 
      multiplexing increases (this corresponds to supposing that as we 
      add channels they do not become worse). Then the above result 
      shows that the ratio of the probabilities of errors to no errors 
      in the useful output channel decreases as $1/T$, ie. error 
      filtration works equally well 
      when the transmission channels are not all identical.

      An interesting question concerns how our methods scale. There are 
      are a number of different issues to be considered. For example, 
      (a) for fixed length $L$ how the intensity of the 
      signal changes when we improve 
      fidelity; (b) for a fixed output fidelity, how the signal 
      changes when we increase $L$; (c) for fixed output fidelity, how 
      the resources required for multiplexing increase as we increase 
      $L$.

      (a) For a fixed length of the communication channel, when we 
      increase the multiplexing in order to increase the fidelity to 1, 
      the total output signal decreases towards a fixed value that 
      depends on the quality of the transmission channel. This value is 
      nothing else than the probability that the signal is not affected 
      by noise in the original, non-multiplexed channel. In other words, 
      as we increase the filtering power of our method (by increasing 
      the multiplexing factor) we identify the errors better and better 
      and throw away a larger fraction of them. Eventually (in the limit 
      of infinitely many multiplexing channels), a perfect filter will 
      yield a perfectly clean signal, (Fidelity =1) while throwing away 
      all the errors, but not more than that. 
      (For example, if the noise is phase noise, as in equation 
      (\ref{phase-noise}), then the probability of an error not to occur 
      is $|\alpha|^2$. In the limit that the Fidelity becomes one, the 
      probability of receiving a signal tends to $|\alpha|^2$.) 

      (b) Each unit of length has equal probability of producing an 
      error. Thus the probability for a signal to survive without being 
      affected by noise (and hence to pass our filtration) decreases 
      exponentially with $L$. This is an inevitable feature of an error 
      filtering method as opposed to one in which errors are corrected. 
      In effect, what our procedure does is to transform a general error 
      (phase noise, depolarization, etc..) into an erasure, which well known to 
      be a considerable advantage.

      (c) The resources required to achieve a fixed fidelity as the 
      length $L$ increases grow polynomially in $L$. This polynomial 
      scaling is achieved by applying error filtration in ``series''. 
      By this we mean that the signal is encoded and after a short distance 
      decoded and the noise filtered out. The signal is then re-encoded 
      and re-decoded many times until the end of the communication 
      channel is reached. Suppose that the error rate per unit distance 
      is $\gamma$ so that the probability that the signal is unaffected 
      by error is $|\alpha|^2= e^{-\gamma L}$. Suppose that the 
      communication channel has length $L$, that the signal is encoded 
      and decoded a total of $Q$ times and that the degree of 
      multiplexing is $T$. Then using 
      eq. (\ref{three3}) one can show that the probability that 
      the signal appears at the useful output and is affected by error is (see 
      the Appendix) \beqa \left[e^{-\gamma L \over Q} + {1-e^{-\gamma 
      L \over Q }\over 
      T} \right]^{Q }-e^{-\gamma L }. \label{Q} \eeqa On the other hand, 
      the probability that the signal appears at the useful output and 
      is not affected by error is $e^{-\gamma L}$. Hence one easily 
      deduces from this result that one can 
      maintain a fixed desired high fidelity of the 
      output signal as we increase $L$ 
      by increasing both the multiplexing factor $T$ and the number of 
      filtering units $Q$ linearly with $L$. 
      In other words the resources required to maintain a fixed fidelity of 
      output signal scale polynomially with the length of the channel. 
      (The need to repeat the filtration step is 
      very similar to existing error correction methods where one also 
      needs to perform the correction step many times, each time before 
      the error probability becomes too large. If the correction - or in 
      our case filtering - is not performed a number of times but only 
      once, then the resources needed to obtain a high fidelity increase 
      exponentially with $L$.)

      Note however that, important as it is, asymptotic scaling is 
      not always the most relevant issue in practice. In practice one 
      always deals with fixed range of distances, and the main question 
      is what is the advantage that a given method yields for that 
      range. This is determined by the asymptotic formula but also by 
      the precise values of the relevant parameters. 
      Thus for instance the BB84 quantum cryptography protocol 
      becomes insecure when the 
      fidelity of the communication channel is below 85\%, and a 
      singlet state becomes an unentangled Werner density matrix when 
      its fidelity is below 50\%. These figures show that 
      degradation of signal with serious consequences can occur 
      with relatively low levels of noise. With a modest level of 
      multiplexing, our schemes can reduce the noise in 
      communication so as to bring states above these important 
      thresholds.

   \section{Error filtration for transmission of entangled states}

The previous protocols illustrated error-filtration for 
      one-way communication of quantum signals. We now 
      show that these ideas may be extended to provide protocols when 
      the goal is to send some {\it known} entangled quantum state from 
      a source to two or more parties. Firstly it is obvious that this 
      may be achieved, and the errors diminished, by using the previous 
      quantum protocols: one of the parties simply prepares the state 
      locally and transmits it using the one-way communication channels 
      we have described earlier. 
      However since the task is to distribute {\it known} entangled 
      states, i.e. it is us that produce them, not some external 
      process out of our control, it is possible to proceed differently 
      by modifying not only the transmission channels but also the 
      source that produces the state. Our method essentially calls for 
      using a source that produces the same number of entangled 
      particles as the original, but in states with more entanglement 
      than what is ultimately needed. We call this method ``source 
      multiplexing''. It provides a new method for {\it 
      entanglement purification}.

      Consider the case in which the aim is for two receivers, $A$ and 
      $B$, to share a quantum state $\rho_m$ which is as close as 
      possible to the maximally entangled state of dimension $m$: $ 
      |\psi_m^R\rangle={1\over \sqrt{m}}\sum_{j=1}^m 
      |j\rangle_A\otimes|j\rangle_B$. To improve the fidelity we use a 
      source that produces two entangled particles in the maximally 
      entangled state of dimension $n$ 
      \begin{eqnarray} 
      |\psi_n\rangle={1\over \sqrt{n}}\sum_{j=1}^n 
      |j\rangle_A\otimes|j\rangle_B \label{one}\ 
      \end{eqnarray} where $n\geq m$, i.e. we start with more 
      entanglement than we want to end up with.

      Again, for simplicity, we consider the case that 
      the dominant errors are of the form (phase noise): 
      \begin{eqnarray} 
      |j\rangle_A |e_0\rangle_A \mapsto |j\rangle_A(\alpha|e_0\rangle_A 
      + \beta|e_j\rangle_A),\label{errors} 
      \end{eqnarray} 
      where $|e_0\rangle$ and $|e_j\rangle$ are states of the 
      environment; $\alpha$ and $\beta$ are complex number satisfying 
      $|\alpha|^2 +|\beta|^2 = 1$. The dominant errors for party $B$ 
      are also of this form. For different states of the system the 
      disturbed states of the environment are orthogonal (${}_A\langle 
      e_j |e_k\rangle_A = {}_B\langle e_j |e_k\rangle_B=0$, $j,k=1...n$, 
      $j\neq k$; ${}_A\langle e_j |e_k\rangle_B=0$ $\forall j,k$). 
      $\alpha$ and $\beta$ describe the amount of disturbance; 
      for simplicity we have taken them to be independent of $j$.

      One may easily show that after going through the noisy channel, 
      the reduced density matrix of the two particles becomes 
      \begin{eqnarray} 
      \rho_n = p P_{\psi_n} + {1 -p \over n} \sum_{j=1}^n P_{jj} 
      \label{five} 
      \end{eqnarray} 
      where $P_{\psi_n} = |\psi_n\rangle\langle\psi_n|$, 
      $P_{jj}=|jj\rangle \langle jj|$ and $p = |\alpha|^4$ is the 
      probability that the state is not affected by phase noise. 

      Let us suppose that the parties do not carry out error 
      filtration. Then the dimensions $n$ and $m$ of the senders and 
      receivers state are equal. We characterise the quality of the 
      receivers state by the fidelity $F_m$, that is the overlap of the 
      receivers state with the state unaffected by noise. 
      One finds that 
      \begin{eqnarray} 
      F_m = {\rm Tr}(\rho_m P_{\psi_m}) = p + {1-p\over m}. 
      \end{eqnarray} 

      Consider now that the source is multiplexed, i.e. $n>m$. The two 
      receivers first carry out a unitary decoding operation: \beqa 
      U_d^A |j\rangle_A &=& {1\over \sqrt{n}} \sum_k e^{-i 2 \pi j k /n} 
      |k\rangle_A \nonumber\\ U_d^B |j\rangle_B &=& {1\over \sqrt{n}} 
      \sum_k e^{+i 2 \pi j k /n} |k\rangle_A \nonumber \eeqa where we 
      have taken $U_d^A$, $U_d^B$ to be 
      the Fourier transform and the inverse 
      Fourier transform respectively. Party $A$ then measures the 
      operators 
      $Q^A = \sum_{k=1}^m |k\rangle_A{}_A\langle k|$, and party $B$ 
      measures the operator 
      $Q^B = \sum_{k=1}^m |k\rangle_B{}_B\langle k|$. By measuring 
      $Q^{A(B)}$ we mean that the party keeps the particle if it is in 
      channels $1$ to $m$, and discards the particle otherwise. It is 
      important that this is a measurement that does not affect the 
      state of the particle if the measurement succeeds. Generalisations 
      of this scheme to other decoding operations and measurements are 
      described in the Appendix. 

      If both measurements succeed, the state becomes 
      \begin{eqnarray} \rho_f &=& \frac{ Q^A Q^B 
      U_d^A U_d^B 
      \rho_n U_d^{A\dagger} U_d^{B\dagger} Q^A Q^B} 
      {Tr (Q^A 
      Q^B U_d^A U_d^B 
      \rho_n U_d^{A\dagger} U_d^{B\dagger})}\nonumber\\ 
      &=& \frac{1}{P_{success}} 
      \big[ \frac{mp}{n}P_{\psi'_m} + \frac{1-p}{n^2} \nonumber\\ 
      &&\sum_{k,k',l,l'=1}^m \delta(k - k' - l + l') |k\rangle_A 
      |l\rangle_{B\ A}\langle k'|_B\langle l'| \big] \nonumber 
      \eeqa 
      where 
      $|\psi'_m\rangle = \frac{1}{\sqrt{m}}\sum_{k=1}^m |k\rangle_A 
      |k\rangle_B$ and $\delta (k - k' - l + l')$ equals $0$ 
      except if $ k - k' - l + l'=0$ mod $n$ when it is equal to 
      $1$; $P_{success}$ is the 
      probability that 
      both measurements succeed and equals 
      \begin{eqnarray}P_{success}= Tr(Q^A Q^B \ U_d^A U_d^B 
      \rho_n U_d^{A\dagger} U_d^{B\dagger} ) = p\frac{ 
      m}{n} + (1-p)\frac{ m^2}{n^2} . \end{eqnarray} 

      The fidelity of the state $ \rho_f$ is 
      \begin{eqnarray} F'_m = Tr( 
      \rho_f P_{\psi'_m}) = \frac{ (n-1)p + 1}{(n-m)p +m} \ . 
      \end{eqnarray} 
      This fidelity is 
      greater than $F_m$, the fidelity if an entangled state of dimension 
      $m$ had been transmitted without error filtration, 
      and tends to one for large $n$. 

      It is important to note that the state will also be purified if 
      party $A$ projects onto the subspace 
      $Q^A_c=\sum_{k=c}^{c+m}P^A_{k}$ and party $B$ simultaneously 
      projects onto the subspace $Q^B_c=\sum_{k=c}^{c+m}P^B_{k}$ for 
      arbitrary $c$. There are $[n/m]$ such orthogonal projectors (where 
      $[x]$ denotes the largest integer smaller or equal to $x$). Thus 
      the total probability that the purification succeeds is 
      $P_{success-total} = [n/m] P_{success}$. We note that, as $n$ 
      becomes large, with $m$ fixed, this total success probability 
      tends to $p$, the probability that no error occurred; in other 
      words we succeed in filtering out all errors.

      In the above procedure the biggest experimental difficulty is 
      apparently the measurement of $Q^A$ and $Q^B$. However in many 
      applications (for instance quantum cryptography) this measurement 
      is not necessary. Indeed the parties may proceed as follows: they 
      assume that the filtration has succeeded and carry out the 
      operations they desire as if the particle is present. After these 
      operations the parties carry out a destructive measurement to 
      check whether the particle is indeed present. If it is then they 
      know that the filtration had succeeded.

      The above ideas are illustrated in Figs. 3 and 4 for photons 
      travelling in multiple fibers and photons travelling in multiple 
      time bins in the same fiber. Note that the protocols in Figs. 3 
      and 4 do not use the Fourier transform as decoding operation, but 
      rather they use a Hadamard transformation. However it is easy to 
      show, (see the Appendix), that these method are both
      effective.  

 \section{Conclusion}  
      In summmary we have presented a 
      conceptually new way of dealing with errors in quantum 
      communication. This idea, error filtration, is a method 
      for reducing errors in quantum communication which can be easily 
      implemented using present day technology. Indeed, to our 
      knowledge it is the first 
      method that can easily be implemented in practice today. For 
      this reason we believe it will find a wide range of applications in 
      quantum 
      information processing and communication.

      \acknowledgments{We gratefully acknowledge partial financial support by 
      the European IST project RESQ, by the Swiss NCCR ``Quantum 
      Photonics'', by the Communaut\'e Fran{\c{c}}aise de Belgique 
      under grant No. ARC 00/05-251, the IUAP program of the Belgian 
      government under grant No. V-18. S.M. is supported by the Belgian 
      National Research Foundation (FNRS).} 

\appendix
\section{Summary of the Appendices}
 
      We extend the protocols presented in the main text in a number of 
      ways. Specifically we consider protocols with more general 
      encoding and decoding 
      operations, protocols where the particle has internal degrees of 
      freedom, protocols in which error filtration is used in series, error 
      filtration for classical wave signals, and general 
      protocols for the transmission of entangled states.

      These results are 
      organised as follows: 
      \begin{itemize} 
      \item {\bf Appendix \ref{sec-phase-noise-general}:
General protocol for error filtration.} 
\\
      The protocol described in the main body of the article took each 
      source channel and multiplexed it into $T$ transmission channels. The 
      encoding and decoding transformations were 
      the Fourier and inverse Fourier transforms respectively. Here we 
      generalise this protocol to other encoding and decoding 
      transformations, and derive a condition for the encoding/decoding to 
      remove as much noise as possible. 
      \item {\bf Appendix \ref{sec-phase-noise-more-general}: 
Protocol for error filtration 
      with collective encoding.}
\\ 
      In the previous protocols a given source channel is encoded into a 
      sub-set of the transmission channels, but the signal carried in a 
      given transmission channel came from a single source channel. Here we 
      show how it 
      is possible to generalise these ideas by allowing each 
      transmission channel to carry signals from more than source 
      channel. 
      \item {\bf Appendix \ref{sec-quantum-internal}:
Error filtration for particles with internal degrees of 
      freedom.} 
\\
      All 
      the previous protocols can be generalised to the case where the particle 
      has ''internal'' degrees of freedom. By this we mean that each source 
      and transmission channel has a state space which is a Hilbert space of 
      dimension larger than one. 
      \item {\bf Appendix \ref{sec-quantum-series}: 
Using error filtration in series.} 
\\     
 The previous protocols can loosely be described as using 
      transmission channels in parallel to achieve noise filtration. We 
      may also use the idea of multiple channels in series to filter 
      noise. By this we mean that the signal is encoded and after 
      transmission over a short distance it is decoded and the noise 
      filtered out. The signal is then reencoded and redecoded several times 
      until one reaches the end of the transmission channel. The final 
      amount of noise at the receiver is less than if there was a single 
      encoding and decoding operation. 
      \item{\bf Appendix \ref{sec-multi-excitation}: 
Quantum multi-excitation protocol.} 
\\  
    In the previous protocols the channels contained only a single 
      excitation each. In the case of bosons, we can also consider the 
      situation where each channel contains many quanta. 
      We show that error filtration also works in this case. We first 
      illustrate this 
      by considering the case that the channel states are coherent 
      states; at the end of this appendix we show how the protocol may be 
      used for general multi-excitation states. 
      \item 
      {\bf Appendix \ref{sec-multi-excitation2}: 
Error filtration for classical wave signals.}
\\
 A particular 
      case of the multi excitation protocol considered in the previous 
      appendix is the case where the number of quanta is macroscopic and 
      one is dealing with classical wave signals. We show that error 
      filtration can also be used to reduce noise when transmitting 
      classical wave signals. In addition to noise precisely analogous 
      to that in previous appendices (phase noise, noise affecting 
internal degrees of 
      freedom) error filtration can also filter out other types of 
      noise such as amplitude noise (i.e. noise that affects the 
      amplitude of the wave) or even non linear noise (when the amount 
      of noise depends on the intensity of the signal). We illustrate 
      how error filtration works for these other types of noise in the 
      case of classical signals. 
      \item {\bf Appendix \ref{sec-quantum-entangled}: 
Protocols for communication of entangled 
      states.} 
\\
      In the main text we described a simple protocol 
      showing 
      how error filtration 
      could be used to filter out noise during transmission of entangled 
      states. Here we generalise this protocol. Two approaches are 
      illustrated. First one may consider a source that produces 
      entangled states of dimension $S$. The entangled particles 
      are then sent to the two receivers, using $T=S$ transmission channels, 
      who project their state onto a smaller Hilbert 
      space of dimension $R$. Here we generalise this protocol and give 
      conditions on the decoding measurements for the protocol to filter 
      out as much noise as possible. Second one may use a source which 
      produces entangled states with the same dimension $S$ as the 
      receiver Hilbert space $R=S$. The number of transmission channels 
      $T$ is taken to be larger than $S$. Thus one is essentially using the 
      protocols developed for error filtration in the case of a single 
      particle twice, once for each particle. 
      We briefly describe how such a protocol filters out errors in the case of 
      entangled particles. 
      \end{itemize}

      \section{General protocol for error 
      filtration}\label{sec-phase-noise-general} 

      In the main part of the paper a specific protocol for error 
      filtration was presented. Here we show how it can be generalised. 
      The generalization consists in allowing more general encoding and 
      decoding operations than the Fourier transform. Nevertheless here 
      we keep the restriction that each source channel is encoded and 
      decoded separately. Therefore we can focus on a particular one, 
      $\ket 1S$, say. 

      We start with the state of the source 
      $\ket 1S$ but now encode the source state into $T$ transmission 
      channels: 
      \beqa \ket 1S \mapsto U_e\ket 1S, \eeqa where the 
      encoding transformation $U_e$ is a unitary map from the source 
      Hilbert space to the transmission Hilbert space. 
      Note that since the transmission Hilbert space is of dimension $T$, 
      whereas there is a single source state, we have to suppose that there 
      are other input channels on which $U_e$ can act. These additional input 
      channels contain no excitations. 
      Thus $U_e^\dagger U_e = (id)_S$ and $U_e U_e^\dagger = (id)_T$, where 
      $(id)_S$ is the 
      identity operator in the source Hilbert space. 

      Following encoding, the state of the system plus environment 
      can be written as \beqa\sum_{j=1}^T \bra jT U_e \ket 1S \quad \ket 
      jT \ket 0E 
      \eeqa 
      where we have introduced an orthonormal basis of transmission states 
      $\ket jT$. 

      The noise now occurs, 
      causing the state to change to \beqa \sum_{j=1}^T \bra jT U_e \ket 
      1S \quad \ket jT \left( \alpha \ket 0E + \beta \ket jE\right) 
      \eeqa 
      where we have supposed that the noise acts independently on each 
      transmission state. 

      We now decode by performing a second unitary 
      transformation (the generalisation of the inverse Fourier 
      transform): \beqa \ket jT \mapsto U_d \ket 
      jT . \eeqa Unitarity means that $U_d^\dagger U_d = (id)_T$ and 
      $U_d U_d^\dagger = (id)_R$. Thus the 
      state becomes \beqa \sum_{j=1}^T \bra jT U_e \ket 1S \quad U_d 
      \ket jT \left( \alpha \ket 0E + \beta \ket jE\right). \eeqa 

      Up to this point the encoding and decoding procedure is very 
      general. We will now restrict ourselves by demanding that if there 
      is no noise, the state should be transmitted exactly into the 
      receiver channel $\ket 1R$. Thus we specify \beqa U_d U_e \ket 1S 
      = \ket 1R.\label{UdUe} \eeqa 

      Thus the state of the system and environment becomes 
      \beqa \ket 1R 
      \left( \alpha \ket 0E + \beta \sum_{j=1}^T \bra jT U_e \ket 1S 
      \bra 1R U_d \ket jT\ket jE 
      \right)+ \beta \sum_{j=1}^T \sum_{k=2}^R c_{kj} 
      \ket kR \ket jE, \label{gen-output-state} \eeqa 
      where 
      \beqa c_{kj} = 
      \bra jT U_e \ket 1S 
      \quad \bra kR U_d \ket jT 
      \eeqa 

      Again, we see that this state has a component in the receiver 
      channel $\ket 1R$, which we regard as the ``useful'' signal and 
      components in all the other receiver channels which we will discard. 

      The state in the useful receiver channel is \beqa \ket 1R \left( 
      \alpha \ket 0E + \beta \sum_{j=1}^T \bra jT U_e \ket 1S \quad \bra 
      1R U_d \ket jT\ket jE \right)\label{gen-useful-state}.\eeqa 

      The probability that the particle ends in the useful channel is 
      the magnitude squared of the state (\ref{gen-useful-state}), i.e. 
      \beqa |\alpha|^2+ 
      |\beta|^2 \sum_{j=1}^T \big|\bra jT U_e \ket 1S \quad \bra 1R U_d \ket 
      jT\big|^2 
      \eeqa 

      On the other hand, there is a probability of 
      \beq 1-(|\alpha|^2 
      +|\beta|^2 \sum_{j=1}^T \big|\bra jT U_e \ket 1S \quad \bra 1R U_d 
      \ket 
      jT\big|^2 )\eeq 
      that the particle appears at one of the ``non-useful" receivers 
      channels, $\ket kR $, $k=2,...T$. These channels are non-useful 
      because the particle ending here is always correlated with noise 
      in the environment. Indeed, we see in (\ref{gen-output-state}) 
      that the component containing the receiver channels $\ket kR $, 
      $k=2,...,T$ is 
      \beqa 
      \beta \sum_{j=1}^T \sum_{k=2}^T c_{kj} \ket kR \ket jE, 
      \eeqa which has no overlap with the unperturbed state of the 
      environment $ \ket 0E$.

      The probability of noise in the useful receiver channel is \beqa 
      |\beta|^2 \sum_{j=1}^T \big|\bra jT U_e \ket 1S \quad \bra 1R U_d \ket 
      jT\big|^2 
      . \eeqa Now Schwartz's inequality shows that this magnitude is 
      greater than 
      \beqa { |\beta|^2\over T} \left|\sum_{j=1}^T \bra jT U_e 
      \ket 1S \quad \bra 1R U_d \ket 
      jT \right|^2= { |\beta|^2\over T} 
      , \eeqa with equality when 
      \beqa 
      |\bra jT U_e \ket 1S \quad \bra 1R U_d \ket 
      jT | = {1\over T} 
      ,\qquad {\rm independent\ of}\ j.\label{independencecondition} 
      \eeqa 
      Thus for any encoding and decoding scheme satisfying the 
      conditions (\ref{UdUe}) and (\ref{independencecondition}), we find 
      that the noise amplitude is reduced by a factor of $1\over \sqrt 
      T$. 

      One example of encoding/decoding schemes satisfying these conditions 
      is the Fourier transform in the previous appendix. A second 
      example is the Hadamard transform; in the case of four 
      transmission channels encoding each source channel, the encoding 
      is 
      \beqa 
      \ket 1S &\mapsto& {1\over 2} ( \ket 1T + \ket 2T +\ket 3T + \ket 4T 
      )\nonumber\\ 
      \ket 2S &\mapsto& {1\over 2} ( \ket 1T + \ket 2T -\ket 3T - \ket 4T 
      )\nonumber\\ 
      \ket 3S &\mapsto& {1\over 2} ( \ket 1T - \ket 2T +\ket 3T - \ket 4T 
      )\nonumber\\ 
      \ket 4S &\mapsto& {1\over 2} ( \ket 1T - \ket 2T -\ket 3T + \ket 4T 
      ); 
      \eeqa 
      the decoding step is 
      \beqa 
      \ket 1T &\mapsto& {1\over 2} ( \ket 1R + \ket 2R +\ket 3R + \ket 
      4R 
      )\nonumber\\ 
      \ket 2T &\mapsto& {1\over 2} ( \ket 1R + \ket 2R -\ket 3R - \ket 
      4R 
      )\nonumber\\ 
      \ket 3T &\mapsto& {1\over 2} ( \ket 1R - \ket 2R +\ket 3R - \ket 
      4R 
      )\nonumber\\ 
      \ket 4T &\mapsto& {1\over 2} ( \ket 1R - \ket 2R -\ket 3R+ \ket 
      4R 
      ). 
      \eeqa

      \section{Protocol for error filtration 
      with collective encoding}\label{sec-phase-noise-more-general} 

      In the previous appendices a given source channel is encoded into a 
      sub-set of the transmission channels, but the signal carried in a 
      given transmission channel came from a single source channel. It 
      is possible to generalise these ideas by allowing each 
      transmission channel to carry signals from more than one source 
      channel. This will be of use, for example, when the number of 
      source channels does not divide the number of transmission 
      channels (eg 2 source channels and 3 transmission channels). 

      Thus the general situation is that we have $S_{tot}$ source 
      channels which we encode collectively into $T_{tot}$ transmission 
      channels and then decode the transmission channels collectively 
      into $R_{tot}$ receiver channels with $S_{tot}=R_{tot}$. 
      It is clear that the previous 
      protocols, in which a given source channel is encoded into $T$ 
      transmission channels, and each source channel is encoded into a 
      different set of transmission channels, is included in this 
      general framework. But other possibilities exist. 

      Here we will simply illustrate the idea with an example. We 
      consider again the case of phase noise: errors in the transmission 
      channels of the form \beqa \ket kT \ket 0E \mapsto \ket kT (\alpha 
      \ket 0E + \beta \ket kE )\quad k=1...T_{tot}. \eeqa If each source 
      channel $\ket jS$ $j=1...S_{tot}$ is simply sent along a single 
      transmission channel (trivial encoding) the error amplitude is 
      $\beta$. However an example of the general framework in the 
      previous paragraph is the following encoding/decoding scheme, 
      based on the Fourier transform. The encoding step is \beqa \ket 
      jS \mapsto {1\over \sqrt T} \sum_{k=1}^{T_{tot}} e^{-2\pi i 
      k(j-1)/T}\ket kT. \eeqa The decoding step is \beqa \ket kT \mapsto 
      {1\over \sqrt T} \sum_{m=1}^{S_{tot}} e^{2\pi i k(m-1)/T}\ket mR. 
      \eeqa 

      For example this protocol may be used in the case that $S_{tot} = 
      2=R_{tot}$ and $T_{tot} = 
      3$. Let us write the state emitted by the source as 
      \beqa 
      a_1 \ket 1S + a_2 \ket 2S 
      \eeqa 
      where $a_1$ and $a_2$ are complex amplitudes obeying $|a_1|^2 + 
      |a_2|^2 = 1$. 
      It may be calculated that the fidelity of the state arriving at the 
      receiver in the channels $\ket 1R$ and $\ket 2R$ to the incoming 
      state is 
      \beqa 
      {|\alpha|^2 + {|\beta|^2\over 3}(1 + 2|a_1|^2|a_2|^2) 
      \over 
      |\alpha|^2 + {2|\beta|^2\over 3}}. 
      \eeqa 
      The average value of this fidelity over the Bloch sphere of 
      incoming states is 
      \beqa 
      {|\alpha|^2 + {4|\beta|^2\over 9} 
      \over 
      |\alpha|^2 + {2|\beta|^2\over 3}}. 
      \eeqa 
      For any $\alpha$ and $\beta$, this is greater than the average fidelity 
      achieved by simply 
      sending each source state through one transmission channel.

      \section{Error filtration for particles with internal degrees of 
      freedom}\label{sec-quantum-internal} 

      The fact that the noise corrected by the previous protocols was 
      phase-noise, and hence corresponds to random elements of the 
      Abelian group $U(1)$ was not critical. The protocols can be simply 
      extended to the case where each channel can carry a system which 
      has ``internal'' degrees of freedom (i.e. each channel 
      has a state space which is a Hilbert space of arbitrary dimension $I$; 
      we shall consider this dimension to be finite here, but this is 
      not essential to the success of the protocol). 

      Thus we consider an orthonormal set of source states 
      \beqa 
      \ket {i\mu} S; \quad \quad i = 1 ... S_{tot},\quad \mu = 1 ... I. 
      \eeqa 
      An example is the case where each channel can carry a spin degree 
      of freedom, so that $I=2$. 
      We consider a set of transmission states 
      \beqa 
      \ket {j\mu} T; \quad j=1...T_{tot},\quad \mu = 1 ... I. 
      \eeqa 
      i.e. there are $T_{tot}$ transmission channels each of which carries a 
      state space of dimension $I$. 
      The transmission states are affected by the following 
      noise: 
      \beqa 
      \ket {j\mu} T \ket 0E \mapsto \alpha \ket {j\mu} T \ket 0E + 
      \sum_{\nu=1}^I \sum_{\lambda=1}^L \beta_\lambda 
      (E_\lambda)_{\mu\nu}\ket 
      {j\nu} T \ket {j\lambda} 
      E.\label{nonAbeliannoise} 
      \eeqa 
      This describes $L$ types of error; 
      each error corresponds to a rotation of the system state. In the case 
      of internal spin degrees of freedom, $I=2$, an example of a set of 
      possible errors is the set 
      of three Pauli matrices: 
      \beqa 
      (E_1)_{\mu\nu}= (\sigma_x)_{\mu\nu};\quad (E_2)_{\mu\nu}= 
      (\sigma_y)_{\mu\nu};\quad (E_3)_{\mu\nu}= (\sigma_z)_{\mu\nu}. 
      \eeqa 
      Thus $ \ket{j \lambda}E $ is the state of the environment if error 
      $\lambda$ occured on channel $j$. 

      The total probability of error in the channel (\ref{nonAbeliannoise}) is 
      \beqa 
      & &\left| \sum_{\nu=1}^I 
      \sum_{\lambda=1}^L \beta_\lambda (E_\lambda)_{\mu\nu}\ket {j\nu} T \ket 
      {j\lambda}E\right|^2 \nonumber \\ 
      &=& \sum_{\lambda=1}^L \left| \beta_\lambda\right|^2 
      \left| \sum_{\nu=1}^I (E_\lambda)_{\mu\nu} \right|^2 
      \nonumber \\ 
      &=& 1 - |\alpha |^2 \label{nonAbeliannoiseprob} 
      \eeqa 
      Thus if the source state $\ket {1\mu}S$ is simply sent through a single 
      transmission channel with trivial encoding and decoding, 
      \beqa 
      \ket {1\mu} S \mapsto \ket {1\mu} T \mapsto \ket {1\mu} R, 
      \eeqa 
      the probability of error is 
      \beqa 
      1 - |\alpha |^2. 
      \eeqa 
      Now consider a simple error-filtration protocol in which the 
      source channel is multiplexed to $T$ transmission channels. The 
      encoding step is 
      \beqa 
      \ket {1\mu} S\ket 0 E \mapsto {1\over \sqrt T} \sum_{j=1}^T \ket 
      {j\mu} T 
      \ket 0E. 
      \label{diag-encod} 
      \eeqa 
      Note that the encoding is independent of $\mu$. 

      Now the noise occurs during the transmission, and the state 
      becomes 
      \beqa 
      & &{1\over \sqrt T} \sum_{j=1}^T \ket {j\mu} T \ket 0E 
      \nonumber\\ 
      & &\quad \mapsto 
      {1\over \sqrt T} \sum_{j=1}^T \alpha \ket {j\mu} T \ket 0E + 
      {1\over \sqrt T} \sum_{j=1}^T\sum_{\nu=1}^I \sum_{\lambda=1}^L 
      \beta_\lambda (E_\lambda)_{\mu\nu}\ket {j\nu} T \ket {j\lambda} 
      E. 
      \eeqa 
      The decoding step for this protocol is also the Fourier transform 
      on the $j$ indices: 
      \beqa 
      \ket {j\nu}T \mapsto {1\over\sqrt{T}} \sum_{k=1}^{T}e^{2i\pi
      j(k-1)/T}\ket{k\nu}R . 
      \eeqa 
      As before we select the term $\ket {1\mu}R$ at the receiver; thus 
      the state in this receiver channel is 
      \beqa 
      \alpha \ket {1\mu} R \ket 0E + 
      {1\over T} \sum_{j=1}^T\sum_{\nu=1}^I \sum_{\lambda=1}^L 
      \beta_\lambda (E_\lambda)_{\mu\nu}\ket {1\nu} T \ket {j\lambda} 
      E. 
      \eeqa 
      Now the probability that the state was affected by noise is the 
      square of the magnitude of the second term: 
      \beqa 
      {1\over T^2} \left | \sum_{j=1}^T\sum_{\nu=1}^I \sum_{\lambda=1}^L 
      \beta_\lambda (E_\lambda)_{\mu\nu}\ket {1\nu} R \ket 
      {j\lambda}E\right |^2 
      = 
      {1\over T} \sum_{\lambda=1}^L |\beta_\lambda|^2\left | \sum_{\nu=1}^I 
      (E_\lambda)_{\mu\nu}\right 
      |^2\label{nonAbeliannoiseprob-multi} 
      \eeqa 
      where we have used the fact that 
      \beqa 
      {}_E\langle j_1 \lambda_1 \ket 
      {j_2\lambda_2}E = \delta_{j_1 j_2} \delta_{\lambda_1 \lambda_2} 
      \quad \mbox{and}\quad {}_R \langle 1{\nu_1}|1{\nu_2}\rangle_R = 
      \delta_{\nu_1 \nu_2} 
      \eeqa 
      Thus comparing (\ref{nonAbeliannoiseprob-multi}) with 
      (\ref{nonAbeliannoiseprob}) we see that the error filtration 
      protocol has reduced the probability of error by a factor of $1/T$, 
      i.e. 
      each error amplitude has been reduced by a factor of $1/\sqrt{T}$. 

      This protocol has essentially used the Fourier transform to encode 
      and decode. It is not difficult to extend the protocol to more 
      general encoding/decodings as was done for phase noise in
      Appendices 
\ref{sec-phase-noise-general}, 
      \ref{sec-phase-noise-more-general}. Furthermore 
      the encoding need not be independent of the internal degrees of freedom 
      $\mu$ (as was the case in eq. (\ref{diag-encod})). 

      \section{Using error filtration in series}\label{sec-quantum-series}

      The previous protocols can loosely be described as using 
      transmission channels in parallel to achieve noise filtration. We 
      may also use the idea of multiple channels in series to filter 
      noise. We illustrate this idea in the case of phase noise.

      We compare two situations. Given a source channel we wish to improve, we 
      can 
      use the encoding described in the main text 
      where we multiplex a single source channel into 
      $T$ transmission channels. This has the effect of causing the 
      noise amplitude to be reduced from $\beta$ to $\beta/\sqrt{T}$, as we 
      showed earlier. If we imagine that the transmission channels have a 
      certain length, $l$, 
      we could perform the same encoding as in the above protocol, but then 
      use the original decoding 
      procedure at the half-way point (or any other point along the 
      transmission channels), then 
      re-perform the encoding, allow the signal to travel for the 
      remaining part of the transmission channel, and finally decode 
      again. As we now show, this protocol gives better error filtration 
      than the protocol without the interior decoding/encoding (assuming 
      that the decoding/encoding module itself does not introduce 
      significant errors). Clearly one could perform the decoding/encoding 
      module at as many interior points as one wishes; we calculate 
      the effect of this below. Thus using error filtration in series 
      is somewhat analogous to the 
      quantum Zeno effect by which evolution is frozen by repeated measurements.

      Recall first that if we do not carry out multiplexing, the state of a 
      particle passing through channel 1 is 
      \beqa 
      \ket 1R \left( \alpha \ket 0E + {\beta} \ket 
      1E\right) , 
      \label{ssss} 
      \eeqa 
      whereas if we multiplex into $T$ transmission channels 
      the state of the system plus environment 
      after transmission is 
      \beqa 
      \ket 1R \left( \alpha \ket 0E + {\beta\over \sqrt T} \ket 
      1E\right). 
      \eeqa 

      Now we imagine decomposing the transmission channel into two 
      halves. We describe the environment Hilbert space as being the 
      tensor product of two Hilbert spaces, one for the first half 
      ($E_1$), and one for the second half ($E_2$) of the transmission. 
      After the first half (in the absence of multiplexing) the state is 
      \beqa \ket 1R \left( \alpha' \ket 0{E1} + {\beta'} \ket 
      1{E1}\right) , \eeqa and after the second half it is \beqa \ket 1R 
      \left( \alpha' \ket 0{E_1} + {\beta'} \ket 1{E_1}\right)\left( 
      \alpha' \ket 0{E_2} + {\beta'} \ket 
      1{E_2}\right).\label{series-halfway2} \eeqa 
      Comparing with eq. 
      (\ref{ssss}), we see that $\alpha'^2 = \alpha$. 

      If we carry out multiplexing in series on the two halves, we find that 
      the state after transmission is 
      \beqa 
      \ket 1R \left( \alpha' \ket 0{E_1} + {\beta'\over \sqrt T} \ket 
      1{E_1}\right)\left( \alpha' \ket 0{E_2} + {\beta'\over \sqrt T} \ket 
      1{E_2}\right).\label{series-halfway} 
      \eeqa

      In order to find 
      out the overall probability for an error to have occurred, we 
      write (\ref{series-halfway}) as 
      \beqa 
      \ket 0R \left( \alpha'' \ket 0{E} + {\beta''} \ket 
      {1''}{E}\right), 
      \eeqa 
      where $\ket{1''}{E}$ is a normalised vector. The probability 
      that the useful receiver state is affected by noise is thus 
      \beqa 
      |{\beta''} |^2 = {(1-|\alpha|)(1+ 2T|\alpha| -|\alpha|)\over T^2}. 
      \eeqa 
      It is not difficult to check that this probability is less than 
      the probability of error without the insertion of the 
      decoding/encoding module (this is equal to $|\beta|^2/T$) for any 
      $\alpha$. 

      More generally one can consider what happens if one has a total of 
      $q$ internal decoding/encoding modules. One gets maximal 
      reduction of error probability when these modules are equally 
      spaced along the transmission channel. In this case the total error 
      probability is found to be 
      \beqa 
      \left[|\alpha|^{2\over(q+1)} + {1-|\alpha|^{2\over(q+1)}\over T} 
      \right]^{(q+1)}-|\alpha|^2. 
      \eeqa 
      We note that this probability tends to 
      \beqa 
      |\alpha|^{2(T-1)\over T} -|\alpha|^2. 
      \eeqa 
      as the number, $q$, of internal decoding/encoding modules tends to 
      infinity.

      \section{Quantum multi-excitation protocol}\label{sec-multi-excitation} 

      In the previous protocols the channels contained only a single 
      excitation each. In the case of bosons, we can also consider the 
      situation where each channel contains many quanta. 
      We will illustrate this first 
      by considering the case that the channel states are coherent 
      states; at the end of this appendix we show how the protocol may be 
      used for general multi-excitation states. 

      Let us consider as before two input channels. Each channel is now 
      described by an infinite dimensional Hilbert space and we may 
      describe the states in terms of the creation and annihilation 
      operators: 
      \beqa 
      [a^1_S ,(a^1_S)^\dagger] = 1;\quad {\rm and}\quad [\tilde a^1_S , 
      (\tilde a^1_S)^\dagger]=1, 
      \eeqa 
      where $a^1_S$ refers to the first channel and $\tilde a^1_S$ to 
      the second. We will work in the Schr\"odinger picture of dynamics. 
      Let the initial state of the system be the following coherent state: 
      \beqa 
      N(\lambda) \exp ({\lambda\over\sqrt 2} ((a^1_S)^\dagger+ e^{i\Phi} 
      (\tilde a^1_S)^\dagger )) 
      \ket 0{sys},\label{coherent} 
      \eeqa 
      where $\ket 0{sys}$ is the vacuum state for the system, and $N(\lambda)$ 
      is a 
      normalisation factor. The phase 
      $\Phi$ allows us to transmit a signal; it will also 
be used later to allow 
      us to measure the effect of the 
      noise. 

      Let us first consider what happens in the absence of filtration, 
      that is, when there is trivial encoding, namely when each source 
      channel evolves into a single transmission channel. The state of 
      the system evolves to \beqa N(\lambda) \exp ({\lambda\over\sqrt 2} 
      ((a^1_T)^\dagger+ e^{i\Phi} (\tilde a^1_T)^\dagger )) \ket 0{sys}. 
      \eeqa The initial state of the environment is a product of 
      states, one for each channel. We denote it $\ket \xi E$. Thus the 
      state of the system plus environment is \beqa N(\lambda) \exp 
      ({\lambda\over\sqrt 2} ((a^1_T)^\dagger+ e^{i\Phi} (\tilde 
      a^1_T)^\dagger )) \ket 0{sys}\ket \xi E. \eeqa The effect of the 
      noise is that there is an interaction between the system and 
      environment. This may be modeled by a unitary transformation of 
      the form \beqa U=\exp i((a^1_T)^\dagger a^1_T B^1 + (\tilde 
      a^1_T)^\dagger \tilde a^1_T\tilde B^1), \eeqa where $B^1$ and 
      $\tilde B^1$ are Hermitian operators acting on the environment 
      Hilbert spaces which we do not need to specify further. 

      Thus after transmission through the noisy channels, the state 
      becomes 
      \beqa 
      N(\lambda) \exp ({\lambda\over\sqrt 2} ((a^1_T)^\dagger e^{iB^1}+ 
      e^{i\Phi}(\tilde a^1_T)^\dagger e^{i\tilde B^1})) 
      \ket 0{sys}\ket \xi E. 
      \label{tmulti} 
      \eeqa 
      We now decode the signal trivially so that the state at the 
      receiver is given by eq. (\ref{tmulti}). 
      Let us now allow these two receiver channels to interfere. This has 
      the effect of transforming the operators 
      $a^1_R$ and $\tilde a^1_R$ into 
      \beqa 
      (a^1_R) \mapsto {1\over\sqrt 2} ( c_R^1 + d_R^1);\quad {\rm 
      and}\quad (\tilde a^1_R) \mapsto {1\over\sqrt 2} ( c_R^1 - 
      d_R^1). 
      \eeqa 
      We now calculate the current in the channel $c_R^1$. This is the 
      expected value of the operator 
      \beqa 
      (c_R^1)^\dagger c_R^1 
      \eeqa in the final state \beqa N(\lambda) \exp ({\lambda\over 2} 
      ((c^1_R)^\dagger (e^{iB^1}+ e^{i\Phi}e^{i\tilde B^1}) + 
      (d^1_R)^\dagger (e^{iB^1}- e^{i\Phi}e^{i\tilde B^1})) \ket 
      0{sys}\ket \xi E. \eeqa This expectation value is \beqa {|\lambda 
      |^2\over 4} \bra \xi E (e^{-iB^1}+ e^{-i\Phi}e^{-i\tilde 
      B^1})(e^{iB^1}+ e^{i\Phi}e^{i\tilde B^1}) \ket \xi E. \eeqa Recall 
      that the state of the environment $\ket \xi E$ is a product of 
      states for the individual channels, thus we may write it as \beqa 
      \ket \xi E = \ket {\xi^1} E\ket {\tilde \xi^1} E. \eeqa Thus for 
      example \beqa \bra \xi E e^{-iB^1} e^{i\tilde B^1} \ket \xi E = 
      \bra {\xi^1}E e^{-iB^1}\ket{\xi^1} E \bra {\tilde \xi^1}E 
      e^{i\tilde B^1}\ket{\tilde \xi^1} E. \eeqa We assume, as in our 
      discussions of the previous protocols, that the noise on different 
      channels is independent, thus we write \beqa 
      \bra {\xi^1}E e^{-iB^1}\ket{\xi^1} E =\alpha^*;\quad 
      \bra {\tilde \xi^1}E e^{i\tilde B^1}\ket{\tilde \xi^1}E=\alpha 
      . 
      \eeqa 
      Therefore the expected value of the current in the channel $c_R^1$ 
      is 
      \beqa 
      {|\lambda |^2\over 2} (1 + |\alpha|^2 \cos \Phi). 
      \eeqa 

      We now consider what happens when 
we encode each of the source channels by 
      multiplexing to $T$ 
      transmission channels. We again start with the coherent state 
      (\ref{coherent}). 
      We illustrate the noise filtration 
      in the case when the encoding is the Fourier transform. This 
      encoding 
      has the effect of transforming the creation operators in the 
      coherent state into 
      \beqa 
      (a^1_S)^\dagger \mapsto {1\over\sqrt T} \sum_{i=1}^T 
      (a^i_T)^\dagger;\quad 
      (\tilde a^1_S)^\dagger \mapsto {1\over\sqrt T} \sum_{i=1}^T 
      (\tilde a^i_T)^\dagger. 
      \eeqa 
      The noise now occurs, causing each creation operator in the 
      coherent state to transform into 
      \beqa 
      (a^i_T)^\dagger \mapsto e^{iB^i}(a^i_T)^\dagger; \quad 
      (\tilde a^i_T)^\dagger \mapsto e^{i\tilde B^i}(\tilde 
      a^i_T)^\dagger. 
      \eeqa 
      We now decode with the inverse Fourier transform, and consider the signal 
      in 
      the two receiver channels defined by the creation operators 
      $(a^1_R)^\dagger $ 
      and $(\tilde a^1_R)^\dagger$. We again allow these to interfere 
      and finally calculate the expected value of the current 
      \beqa 
      (c_R^1)^\dagger c_R^1 
      \eeqa 
      in the final state. This is 
      \beqa 
      & & {|\lambda |^2\over 4T^2} 
      \bra \xi E (\sum_{i=1}^T (e^{-iB^i}+ e^{-i\Phi}e^{-i\tilde B^i})) 
      (\sum_{j=1}^T (e^{iB^j}+ e^{i\Phi}e^{i\tilde 
      B^j})) 
      \ket \xi E 
      \nonumber \\ 
      &= &\qquad {|\lambda |^2 \over 2}({1 + (T-1)|\alpha|^2\over T} ) 
      ( 1 + {T|\alpha|^2\over 1 + 
      (T-1)|\alpha|^2} \cos {\Phi}). 
      \eeqa 
      Exactly as in the previous protocols the multiplexing has the 
      effect of reducing the noise. 

      So far in this appendix we have considered a particularly simple 
      initial state, a coherent state. In this case it is rather 
      straightforward to calculate the effect of our filtration 
      protocol. However the protocol may be used for any 
      multi-excitation state. 

      Let us consider that the state of the source is defined by some function 
      of 
      creation operators $(a^1_S)^\dagger$ acting on the vacuum. The 
      effect of the encoding and decoding that we have performed above 
      is to change this state to one in which the operator 
      $(a^1_S)^\dagger$ is transformed to an expression of the form 
      \beqa 
      {1\over T} \sum_{j=1}^T e^{iB^j}(a^1_R)^\dagger. 
      \eeqa 
      Thus any power of the operator $((a^1_S)^\dagger)^N$ becomes 
      replaced by 
      \beqa 
      {1\over T^N} (\sum_{j=1}^T e^{iB^j})^N 
      ((a^1_R)^\dagger)^N.\label{apowerN} 
      \eeqa 
      We now imagine computing the expectation value of some operator in 
      the state. For $T$ much larger than $N$ we can neglect all terms 
      in the expectation value in which any given operator 
      $e^{iB^k}$, say, appears to any power greater than 1. Hence 
      when we compute the expection value, we can perform the 
      inner-product with the state of the environment, and hence replace 
      (\ref{apowerN}) by 
      \beqa 
      {1\over T^N} (T\alpha)^N 
      ((a^1_R)^\dagger)^N, 
      \eeqa 
      where $\alpha$ is the expected value of $e^{iB^k}$ 
      for channel $k$. Thus in the limit of large $T$ we see that 
      the effect of the protocol is that the source operator $(a^1_S)^\dagger$ 
      gets transformed to 
      \beqa 
      \alpha (a^1_R)^\dagger. 
      \eeqa 
      The key point that this protocol achieves (for large $T$) is that 
      interference between operators is 
      not affected i.e. 
      \beqa 
      {1\over\sqrt 2} ((a^1_S)^\dagger+ e^{i\Phi}(\tilde a^1_S)^\dagger) 
      \mapsto \alpha {1\over\sqrt 2} ((a^1_R)^\dagger+ e^{i\Phi}(\tilde 
      a^1_R)^\dagger). 
      \eeqa 
      Destruction of interference is avoided and replaced by overall 
      absorption of quanta. This is the exact analogue of what happens 
      for the single-quanta protocols presented earlier where visibility 
      is improved at the cost of overall reduction in intensity. 

      We note that while we have focused on the case of large $T$ in the 
      previous paragraph, similar analysis shows that, quite generally, 
      even for finite $T$, 
      multiplexing has the effect of reducing noise and replacing it by 
      an overall reduction in intensity.

      \section{Error filtration for classical wave 
      signals}\label{sec-multi-excitation2}

      In the previous appendix we considered the case where
      the signal
      contained many excitations. A limiting case is the one where the
      number of excitations is
      macroscopic and one is dealing with classical wave signals.
      Classical waves (as well as the quantum multi-excitation states
      discussed in the previous appendix) have a number of properties
      that can all be affected by noise. Scalar classical waves
      (i.e. described by a single complex amplitude) can be affected by
      phase noise; waves having ``internal degrees of freedom" (such as
      polarization) can be affected by noise acting on these degrees of
      freedom. But the wave can also be subjected to noise that affects
      its amplitude (amplitude noise). Furthermore, it is also possible
      that the amount of noise in all the above cases depends
      non-linearly on the the amplitude (``non-linear noise"). The very
      same mulplexing scheme that we used for single particles will
      filter noise in all these situations.

      We consider the simple encoding/decoding in which a single source
      channel is encoded in $T$ transmission channels. Denote by
      $\psi_S$ a normalised mode of the source channel. Consider a
      classical signal of amplitude $A$ emitted by the source in mode
      $\psi_S$. Denote by $\psi_T^j$ ($j=1,\ldots,T$) normalised modes
      of the transmission channels. The encoding operation transforms
      the signal as
\beqa A \psi_S \to \sum_{j=1}^T { A \over
      \sqrt{T}}\psi_T^j = \sum_{j=1}^T A_{T_{in}}^j\psi_T^j
\eeqa
where
      $A_{T_{in}}^j$ is the amplitude at the input of transmission
      channel $j$. Suppose that noise acts during the transmission. Then
      the amplitude in transmission channel $j$ gets transformed as
     \beqa
A_{T_{in}}^j \to A_{T_{out}}^{j}(\xi_j, A_{T_{in}}^j) \eeqa where $\xi_j$
      is a random variable (the noise) and
$A_{T_{out}}^{j}(\xi_j, A_{T_{in}}^j)$ is
      the amplitude at the output of the transmission channel if the
      noise has value $\xi_j$ and the signal at the input of the
      transmission channel  has amplitude $A_{T_{in}}^j$.
It is convenient to rewrite the output amplitude as
\beqa
 A_{T_{out}}^{j}(\xi_j, A_{T_{in}}^j) =  A_{T_{in}}^j N^j(\xi_j, A_{T_{in}}^j)
\eeqa
where
$N^j$ is the noise acting on channel $j$.
The noise is linear if $N^j$
      depends only on $\xi_j$ but not on $A_{T_{in}}^j$.

Again, by suitable engineering, we can
      arrange that the noise acts independently on the different
      channels. Mathematically this is equivalent to the statement that the
$\xi_j$ are independent random
      variables. If in addition the channels are
      identical then we have the further simplification that the
      functions $N^j(\xi_j, A_{T_{in}}^j)=N(\xi_j, A_{T_{in}}^j)$ are
      independent of $j$ and that the
$\xi_j$ are independent identically distributed (i.i.d.) random
      variables. For simplicity we assume from now on
that the noise in all
      transmission channels is identical and independent, ie.
\beqa
 A_{T_{out}}^{j}=  A_{T_{in}}^j N(\xi_j, A_{T_{in}}^j)
\quad \mbox{with $\xi_j$ i.i.d. random variables.}
\eeqa

As an illustration linear phase noise is described by a noise
function \beqa N_{linear\ phase}(\xi,A)=e^{i\varphi(\xi)}\eeqa and
linear amplitude noise is described by a function \beqa N_{linear\
amplitude}(\xi,A)=f(\xi)\quad \mbox{with $f$ real}\eeqa whereas
non linear phase noise could for instance be described by a noise
function \beqa N_{nonlinear\
phase}(\xi,A)=N_1e^{i\varphi(\xi)|A|^2}=N_1N_2\label{nonlinnoise}
,\eeqa where $N_1$ represents some linear phase noise and
$N_2=e^{i\varphi(\xi)|A|^2}$ is independent of $N_1$ and describes
the non-linear part.

      Having described the action of the noise, let  us consider the
     form of the useful signal in the receiver channel. We suppose that
the decoding is realised using the
      Fourier transform.
      Denote by $\psi_R$ a normalised mode of the
      receiver channel.
      The amplitude in the receiver channel is
      \beqa
      A_R \psi_R &=& \sum_j {A^{j}_{T_{out}} \over \sqrt{T}}\psi_R
     \nonumber\\
&=&A \left({1\over T} \sum_{j=1}^T N(\xi_j,{A / \sqrt{T}})\right)\psi_R
\ .
\label{ampRC}
      \eeqa

In order to interpret the expression eq. (\ref{ampRC}) we first
concentrate on the case of linear noise. We will then come back to the case
of non linear noise.
Upon averaging over the random noise variables $\xi_j$, one finds that
     the average amplitude in the receiver channel
\beqa
\overline{ A_R }
=
A\  \overline{ N }
\label{AR}
\eeqa
is independent of the degree $T$ of multiplexing.
On the other hand the average intensity in the reception channel
depends on the degree of multiplexing:
\beqa
\overline{I_R}=\overline{A^*_R A_R}
=
|\overline{A_R}|^2 \left(
1 + \frac{1}{T} \frac{\overline{N^* N} -
  |\overline{N}|^2}{|\overline{N}|^2}\right)\ .
\label{IR}
\eeqa
The average intensity thus decreases with the degree of
multiplexing. This is because the filtration removes more and more
noise as $T$ increases. In the limit of large $T$ the average
intensity is equal to the norm squared of the average amplitude.

This result can be reexpressed in terms of the
amplitude fluctuations in the reception channel. These fluctuations
are given by
\beqa
\Delta A_R^2=\overline{A_R^* A_R} - |\overline{A_R}|^2
=|\overline{A_R}|^2 \frac{1}{T} \frac{\overline{N^* N} -
  |\overline{N}|^2}{|\overline{N}|^2}\ .
\label{DR} \eeqa They decrease with $T$. In the limit of large $T$
the amplitude in the useful receiver channel $A_R$  no longer
fluctuates, ie. all the noise has been removed.

It is interesting to also look at the amplitudes in the non useful
receiver channels. These are given by \beqa A_R^k \psi_R^k =
\sum_j \frac{A^j}{\sqrt{T}}e^{i 2 \pi j k /T}\psi_R^k\quad ,\quad
k\neq 0\ . \label{ARk} \eeqa One finds that the average amplitude
in the non useful receiver channels is zero \beqa
\overline{A_R^k}=0 \label{ARkav} \eeqa and that these channels
contain non zero average intensity. This means that these channels
contain only noise.

As an illustration of the effect of the noise reduction in the
useful receiver channel we consider the visibility of interference
fringes. Suppose the sender prepares two signals \beqa
A_{S1}\psi_{S1} + A_{S2}\psi_{S2} \eeqa where
$A_{S1}={1\over{\sqrt 2}}{\cal A}$ and $A_{S2}={1\over{\sqrt
2}}e^{i\phi}{\cal A}$. Let us suppose that each signal is
independently transmitted by multiplexing it into $T$ channels.
The useful receiver signals are \beqa  A_{R1}\psi_{R1} +
A_{R2}\psi_{R2} \eeqa

where
\beqa A_{R1}&=&\frac {\cal A} {\sqrt 2 T} \sum_{j=1}^T N(\xi_j)\nonumber\\
A_{R2}&=&\frac {e^{i\phi}{\cal A}}{\sqrt 2 T} \sum_{j=T+1}^{2T}
N(\xi_j)\nonumber\\ & &\mbox{$\xi_j$ i.i.d. random variables,
$j=1,\ldots,2T$}\ .\label{AR12} \eeqa If the receiver lets the two
signals interfere he will find an intensity \beqa\label{IR12}
\overline{ |\frac{A_{R1} + A_{R2}}{\sqrt{2}}|^2}&=& {\frac 1
2}|{\cal A}|^2 |\overline N|^2\left( 1 + \frac{\overline{|N|^2} -
|\overline{N}|^2}{T
  |\overline{N}|^2}+\cos \phi\right)
\eeqa hence the visibility of interference fringes he sees is
given by \beqa\label{V12}
V=\frac{I_{max}-I_{min}}{I_{max}+I_{min}}
=
\frac{1}{1 + \frac{\overline{|N|^2} - |\overline{N}|^2}{T
  |\overline{N}|^2}} \ .
\eeqa
The visibility thus tends to 1 as the degree of multiplexing $T$ increases.

Let us now consider the case of non linear noise. The effects of
the multiplexing are more complex. Multiplexing reduces noise via
two independent processes. First, the filtration itself. Second,
multiplexing also reduces the intensity in each transmission
channel, and so it reduces nonlinearity. Both effects are
beneficial, and the exact result depends on the interplay between
them, and on the specific form of the non-linearity.

All the above formulae eqs. (\ref{AR} to \ref{V12}) stay valid,
but one must replace $N(\xi)$ by $N(\xi,A/\sqrt{T})$ which
introduces an additional dependence on $T$. A first consequence is
that the average amplitude $\overline{A_R}$  depends now on the
degree of multiplexing. For the noise considered in
(\ref{nonlinnoise}) $\overline{A_R}={\cal A}{\overline
{N_1}}{\overline {N_2}} $ depends on the degree of multiplexing
through $\overline {N_2} $. Let us now compute the noise
  fluctuations (which appear in
  eqs. (\ref{IR},\ref{DR},\ref{IR12},\ref{V12}))
\beqa \frac 1 T \frac{\overline{|N|^2} - |\overline{N}|^2}{
  |\overline{N}|^2}
\label{DN}. \eeqa

As an illustration consider non linear
      phase noise described in (\ref{nonlinnoise}).
Let us suppose that the noise is small so that we can expand
$N_2(\xi,A)=1+ i\varphi|A|^2 - \varphi^2|A|^4/2$. Then one easily
obtains that

\beqa \frac 1 T \frac{\overline{|N|^2} - |\overline{N}|^2}{
  |\overline{N}|^2}= \frac 1 T \Bigg[\frac{\overline{|N_1|^2} - |\overline{N_1}|^2}{
  |\overline{N_1}|^2}+\frac 1 {T^2}\frac {|{\cal A}|^4}
  {|\overline{N_1}|^2}(\overline{\varphi^2}-{\overline\varphi}^2)+O(\frac
  1 {T^4})\Bigg]
. \eeqa Here we can see the interplay of the two ways in which
noise is reduced by multiplexing. The factor $\frac 1 T$ that
multiplies the whole expression comes directly from filtering,
while the factor $\frac 1 {T^2}$ in the square bracket that
multiplies the non linear noise appears because the reduction in
intensity in each transmission channel has reduced the non
linearity.

 Above we considered the situation in which the
      classical wave was described by a single complex amplitude, i.e.
      there were no internal degrees of freedom. In the case of quantum
      systems, we showed in Appendix C, that multiplexing works for
      equally well if the particle has internal degrees of freedom. It
      is not difficult to check that exactly the same is true for
      classical waves.

      All the above remarks which have been made in the case of classical
      signals are important because they also apply in the case of
      signals with one or many excitations: error filtration will remove
      amplitude noise and non linear noise in the quantum case as well.

      \section{Protocols for communication of entangled 
      states}\label{sec-quantum-entangled} 

      Here we generalise the error filtration protocol for communication of 
      entangled states given in the main part of the paper. 

      The protocols we consider have the following general structure. 
      A source emits two S-level systems in an entangled state. System A 
      will be sent to party A whereas system B will be sent to party 
      B. These systems are first 
      encoded as T-level systems. The signals are then 
      transmitted through the noisy channels. Finally they are decoded by by 
      parties A and B 
      to R-level systems. The key to the performance of the 
      protocols is to allow the encoding and decoding operations to change 
      the dimension of the space of states.

      Denote the states of system $A$ when emitted by the source as 
      \beq \kett iSA \ \qquad 
      (i=1...S). \eeq 
      These states are encoded by the encoder 
      by a transformation $U^A_e$; thus the state of 
      the system which emerges is \beq U_e^A\kett iSA = \sum_{j=1}^T 
      \braa jTA U_e^A\kett iSA \ \kett jTA. \eeq 
      The initial state 
      of the system plus environment of the channel is thus \beq 
      \sum_{j=1}^T \braa jTA U_e^A\kett iSA \ \kett jTA\kett 0EA. 
      \eeq After transmission through the channel the state becomes \beq 
      \sum_{j=1}^T \braa jTA U_e^A\kett iSA \ \kett jTA \big(\alpha_{jA}\kett 
      0EA + 
      \beta_{jA}\kett jEA \big). \eeq The decoding acts by a further 
      transformation $U_d^A$. Thus the final state after encoding and 
      decoding is \beq 
      \sum_{j=1}^T \braa jTA U_e^A\kett iSA \ U_d^A\kett jTA 
      \big(\alpha_{jA} 
      \kett 0EA + \beta_{jA}\kett jEA \big). \eeq 

      Finally, only some of the receiver states are used. This is 
      accomplished by projecting the final state on a Hilbert subspace 
      of receiver states of dimension $R$. We will denote this projector 
      by $\Pi^A$. There is a certain liberty in choosing this projector 
      as one can modify the projector by a unitary transformation which, 
      on the other hand, can be absorbed into the definition of $U_d^A$. 
      For concreteness, we fix this arbitrariness by defining 

      \beq \Pi^A=\sum_{l=1}^R \kett kRA \braa kRA. \eeq 

      Thus when starting with the initial state $\kett iSA \kett 0EA $ 
      we end up with the (unnormalized) state \beq 
      \sum_{j=1}^T \braa jTA U_e^A\kett iSA \ \Pi^AU_d^A\kett jTA 
      \big(\alpha_{jA}\kett 0EA + \beta_{jA}\kett jEA \big). \eeq

      Party B performs similar operations on its state. 

      A given protocol is a choice of the dimensions of the different 
      Hilbert spaces at each time, and of the encoding and decoding 
      operations. Below we give some specific examples of these choices.

      \subsection{Protocol 1: Multiplexing at the source} 
      \label{multiplexing-at-the-source} 

      In this example the task is to share an $R$-dimensional maximally 
      entangled state. The protocol works by preparing an $S$-level 
      system ($S>R$), allowing it to be transmitted through the noisy 
      channel, and then processing it at the end (i.e. the encoding 
      between the source and transmitter channels is trivial and $T=S$). 
      The protocol allows the two parties to end up with a final state 
      of their $R$-level systems which has higher fidelity than would 
      have been achieved if the $R$-level system were simply transmitted 
      directly through the channel. We call this method ``multiplexing 
      at source'' because we use a source which produces more 
      entanglement than the one we wish to produce at the receivers 
      ($S>R$). It is this which enables us to obtain a state at the 
      receivers which is closer to the required state than that we would 
      have obtained had we started with the source simply producing an 
      $R$-dimensional entangled state. 

      Consider the initial state 

      \beq |\psi_{in}\rangle= {1\over \sqrt{S}}\sum_{i=1}^S \kett iSA 
      \kett iSB \kett 0EA \kett 0EB. \eeq The first condition defining 
      this protocol is that the encoding stage is trivial. This means 
      that \beq \braa jTA U_e^A \kett iSA = \braa jTB U_e^B \kett iSB 
      = 
      \delta_{ij} \eeq 

      Thus the un-normalised state of the system after decoding and 
      projection is \beq |\psi_{fin}\rangle={1\over 
      \sqrt{S}}\sum_{i=1}^S\Pi^A\Pi^B U_d^AU_d^B \kett iTA \kett iTB 
      (\alpha \kett 0EA + \beta\kett iEA)(\alpha \kett 0EB + \beta\kett 
      iEB) \eeq 

      We are interested in the maximum fidelity to an $R$-level 
      $|\psi_R\rangle$ singlet that we can produce \beq 
      |\psi_R\rangle={1\over \sqrt{R}}\sum_{i=1}^R \kett iRA \kett iRB . 
      \eeq 

      The fidelity of the state $|\psi_{fin}\rangle$ is \beq 
      F={{|\langle\psi_R|\psi_{fin}\rangle|^2}\over{\langle\psi_{fin}| 
      \psi_{fin}\rangle}} \eeq

      We now introduce the second condition defining the protocol, 
      namely that $U^A_d$ and $U^B_d$ should be related by being 
      essentially the complex conjugates of each other in the bases we 
      are using. i.e. If we write \beq U_d^A \kett iTA =\sum_j 
      u_{ij}\kett jRA \eeq then \beq U_d^B\kett iTB =\sum_j 
      u_{ij}^*\kett jRB. \eeq 
      This means in particular that \beq 
      U_d^AU_d^B \sum_i \kett iTA \kett iTB= \sum_i \kett iRA \kett iRB 
      , \eeq 
      ie. in the absence of noise the receiver obtains a maximally 
      entangled state. 

      Now let us compute \beqa 
      \langle\psi_{fin}|\psi_{fin}\rangle &=& {1\over S}\sum_{i,i'=1}^S 
      \braa {i'}TA\braa {i'}TB 
      (U_d^A)^{\dagger}(U_d^B)^{\dagger}\Pi^A\Pi^B U_d^AU_d^B 
      \kett {i}TA\kett {i}TB\nonumber\\ 
      & &\qquad \times\left[|\alpha|^4+ 
      (1-|\alpha|^4)\delta_{i,i'}\right]\nonumber\\ 
      &=& 
      {|\alpha|^4\over S}\sum_{i,i'=1}^S \braa {i'}TA\braa {i'}TB 
      (U_d^A)^{\dagger}(U_d^B)^{\dagger}\Pi^A\Pi^B U_d^AU_d^B 
      \kett {i}TA\kett {i}TB 
      \nonumber\\ 
      \qquad &+& {1-|\alpha|^4\over S}\sum_{i=1}^S \braa {i}TA\braa 
      {i}TB (U_d^A)^{\dagger}(U_d^B)^{\dagger}\Pi^A\Pi^B U_d^AU_d^B 
      \kett {i}TA\kett {i}TB. \eeqa It may be calculated that \beqa 
      \langle\psi_{fin}|\psi_{fin}\rangle = {|\alpha|^4 R\over S} + 
      {1-|\alpha|^4\over S} \sum_{i=1}^S 
      \left( 
      \braa iTA (U_d^A)^{\dagger}\Pi^A U_d^A \kett iTA 
      \right)^2. 
      \eeqa Where we have used the fact that \beqa \braa iTA 
      (U_d^A)^{\dagger}\Pi^A U_d^A \kett iTA = \braa iTB 
      (U_d^B)^{\dagger}\Pi^B U_d^B \kett iTB \eeqa Also \beqa 
      |\langle\psi_R|\psi_{fin}\rangle|^2 = {|\alpha|^4 R\over S} + 
      {1-|\alpha|^4\over RS} \sum_{i=1}^S 
      \left( 
      \braa iTA (U_d^A)^{\dagger}\Pi^A U_d^A \kett iTA 
      \right)^2. 
      \eeqa 

      Thus we may write the fidelity as \beq 
      F={{|\langle\psi_R|\psi_{fin}\rangle|^2}\over{\langle\psi_{fin}| 
      \psi_{fin}\rangle}} = \left({|\alpha|^4 R\over S} + 
      {1-|\alpha|^4\over S} {Y\over R}\right) 
      \left( {|\alpha|^4 R\over S} + {1-|\alpha|^4\over S} {Y}\right)^{-1} 
      \eeq where \beqa Y = \sum_{i=1}^S 
      \left( 
      \braa iTA (U_d^A)^{\dagger}\Pi^A U_d^A \kett iTA 
      \right)^2. 
      \eeqa $Y$ is a positive quantity and by Schwarz's inequality \beqa 
      Y = \sum_{i=1}^S 
      \left( 
      \braa iTA (U_d^A)^{\dagger}\Pi^A U_d^A \kett iTA 
      \right)^2 \geq 
      {1\over S}\left( \sum_{i=1}^S 
      \braa iTA (U_d^A)^{\dagger}\Pi^A U_d^A \kett iTA 
      \right)^2. 
      \eeqa But \beqa \sum_{i=1}^S \braa iTA (U_d^A)^{\dagger}\Pi^A 
      U_d^A \kett iTA = R. \eeqa Thus \beq Y\geq {R^2\over S} \eeq with 
      equality when \beq \braa iTA (U_d^A)^{\dagger}\Pi^A U_d^A \kett 
      iTA = {R\over S}\quad \mbox{for all $i$.} 
      \label{protocolcondition3} \eeq 

      We will impose (\ref{protocolcondition3}) as the third condition 
      defining this protocol. In this case, the fidelity is \beq F= 
      \left({|\alpha|^4 R\over S} + {1-|\alpha|^4\over S} {R\over 
      S}\right) 
      \left( {|\alpha|^4 R\over S} + (1-|\alpha|^4) ({R\over S})^2\right)^{-1}. 
      \eeq 

      We see that by increasing the amount of entanglement produced by 
      the source, (i.e. by increasing $S$), the fidelity is increased and 
      tends to 1 for large S. 

      \subsection{Protocol 2} 
      We may also use the protocols for error filtration presented in 
      the main text and 
      Appendices \ref{sec-phase-noise-general}, 
      \ref{sec-phase-noise-more-general} 
      directly 
      to filter errors when communicating 
      entangled quantum states. We can think of the protocols in the main 
      text and in 
      Appendices \ref{sec-phase-noise-general}, 
      \ref{sec-phase-noise-more-general} as ways 
      of 
      improving a given transmission channel: 
      by multiplexing each source channel to $T$ transmission channels 
      we can reduce the error amplitude from $\beta$ to $\beta/\sqrt{T}$. 

      Consider, then, that a source prepares a state of two $S$ level 
      systems. This state is pre-processed by multiplexing each source 
      channel into $T$ transmission channels 
      using a general encoding as 
      given in Appendix 
\ref{sec-phase-noise-general}. The signal is then decoded 
      and 
      post-processed to yield a state at the two receivers $R_A$ and 
      $R_B$. The received state will be of higher fidelity than if the 
      pre- and post- processing had not been used. 

      Consider for example the following input state \beq 
      |\psi_{in}\rangle= \sum_{i=1}^S a_i\kett iSA \kett iSB, \eeq where 
      $a_i$ are complex amplitudes (this is essentially the most general 
      bi-partite state). If each source channel is processed through $T$ 
      transmission channels, in such a way that the original error 
      amplitude is reduced from $\beta$ to $\beta/\sqrt{T}$, then the 
      final state is \beqa |\psi_{fin}\rangle = \sum_{i=1}^S a_i\kett 
      iSA \kett iSB (\alpha\kett 0EA + {\beta\over \sqrt T} \kett iEA) 
      (\alpha\kett 0EB + {\beta\over \sqrt T} \kett iEB). \eeqa The 
      fidelity of the state at the receivers to the state which would 
      have been transmitted if there were no noise (i.e. $\sum_{i=1}^S 
      a_i\kett iRA \kett iRB$) is 

      \beqa 
      F&=&{{|\langle\psi_{in}|\psi_{fin}\rangle|^2}\over{\langle\psi_{fin} 
      |\psi_{fin}\rangle}}\nonumber\\ 
      &=& { |\alpha|^4 + \left[ \left(|\alpha|^2 + |\beta|^2/T 
      \right)^2 - |\alpha|^4 \right]\sum_{i=1}^S |a_i|^4 \over 
      \left(|\alpha|^2 + |\beta|^2/T \right)^2} . \eeqa Thus the 
      fidelity increases monotonically with $T$ and tends to 1 as 
      $T\rightarrow \infty$.

    \bigskip
      {\bf{Figure Captions}} 

      \bigskip 
      \noindent 
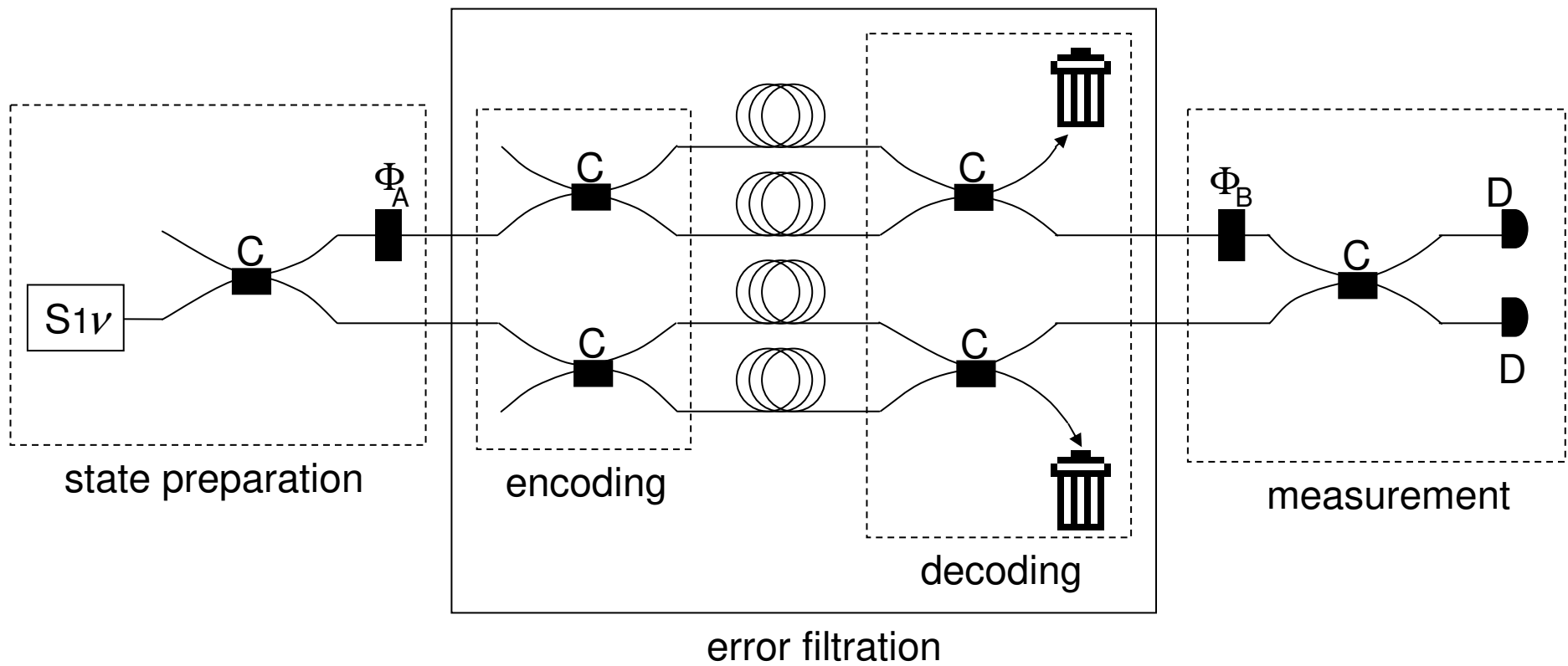
      {\bf Fig. 1.} 
      Implementation of error filtration using multiple optical fibers. 
      A source ($S_{1\nu}$) produces a single photon that is coupled into an 
      optical fiber. The photon is split into two using a fiber coupler 
      (C). Note that an arbitrary state in a two dimensional space (a qubit) 
      can be prepared in this way by changing the 
      coupling ratio of the coupler and by modifying the phase $\phi_A$. 
      In order to protect the state against noise each basis state is 
      multiplexed into two transmission states using 50/50 couplers. A 
      single qubit is thus encoded into 4 transmission states, each 
      traveling through a separate fiber. Because the photon cannot jump 
      from one fiber to the other it will only be affected during 
      transmission by phase noise. Furthermore the noise in the 
      different fibers will be independent. Hence the noise is of the 
      type (independent phase noise on the different transmission 
      channels) studied in the main text. The decoding is the reverse of 
      the encoding procedure: two transmission states are combined into 
      one receiver state using fiber couplers. This is done in such a 
      way that in the absence of noise constructive interference occurs 
      and the photon always emerges in the receiver states. Due to the 
      noise the photon may not emerge in the receiver states in which 
      case it is discarded. But if the photon emerges in the receiver 
      states, then the noise has been filtered out. The receiver can 
      then use the filtered state. For instance he may, as described in 
      the text, test the quality of the filtered state by carrying out 
      the measurement shown (D, single photon detector). The measurement 
      basis is chosen by varying the phase $\phi_B$. Note however that 
      the measurement step is included in the figure for illustration 
      only - it is not part of the filtration protocol per se. The 
      receiver can use the filtered signal for other purposes.

      \bigskip 
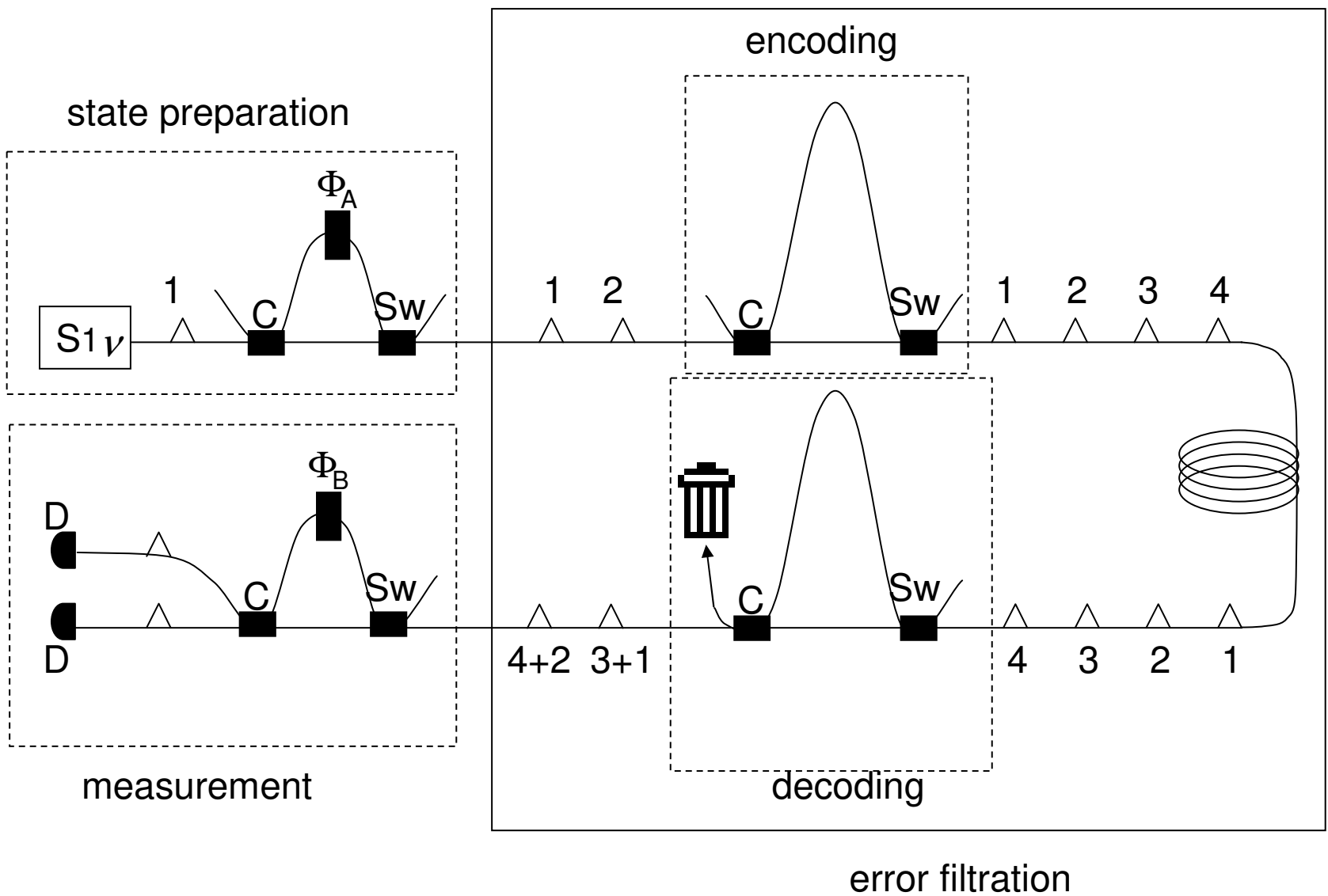
      \noindent {\bf Fig. 2.} Implementation of error filtration using 
      time bins propagating in optical fibers. A source produces a 
      single photon in time bin $|1\rangle$. A first Mach-Zender (MZ) 
      interferometer produces a state in a two dimensional Hilbert space 
      as follows: the fiber coupler C splits the time bin $|1\rangle$ 
      into two pulses which follow the short and long arm of the 
      interferometer. Then the switch (Sw), synchronized with the 
      source, is used to direct the pulses exiting from the first MZ 
      interferometer into the fiber leading to the encoder. In this way 
      a superposition of two time bins $\left( |1\rangle + 
      e^{i\phi_A}|2\rangle\right)/\sqrt{2}$ is produced where the phase 
      $\phi_A$ encodes the quantum information that must be transmitted. 
      A second MZ interferometer realises the encoding part of the error 
      filtration protocol: it multiplexes time bin 1(2) exiting from 
      the state preparation into time bins 1 and 3 (2 and 4). Thus after 
      encoding the qubit is a superposition of 4 time bins $\left( 
      |1\rangle + |3\rangle + e^{i\phi_A} |2\rangle + e^{i\phi_A} 
      |4\rangle \right)/2$. During transmission the state is affected by 
      noise. If the time bins are sufficiently separated, then the 
      probability that a photon jumps from one time bin to the other is 
      negligible and the noise affecting each time bin will be 
      essentially independent. Hence one is in the case of independent 
      phase noise considered in the main text. The decoding operation is 
      the reverse of the encoding operation. A first MZ interferometer 
      projects the state onto the $|1\rangle + |3\rangle$, $|2\rangle + 
      |4\rangle$ subspace. The receiver may then use the filtered state. 
      For instance he can carry out a measurement using a second MZ 
      interferometer. The measurement basis is chosen by varying the 
      phase $\phi_B$. 

      \bigskip 
      \noindent 
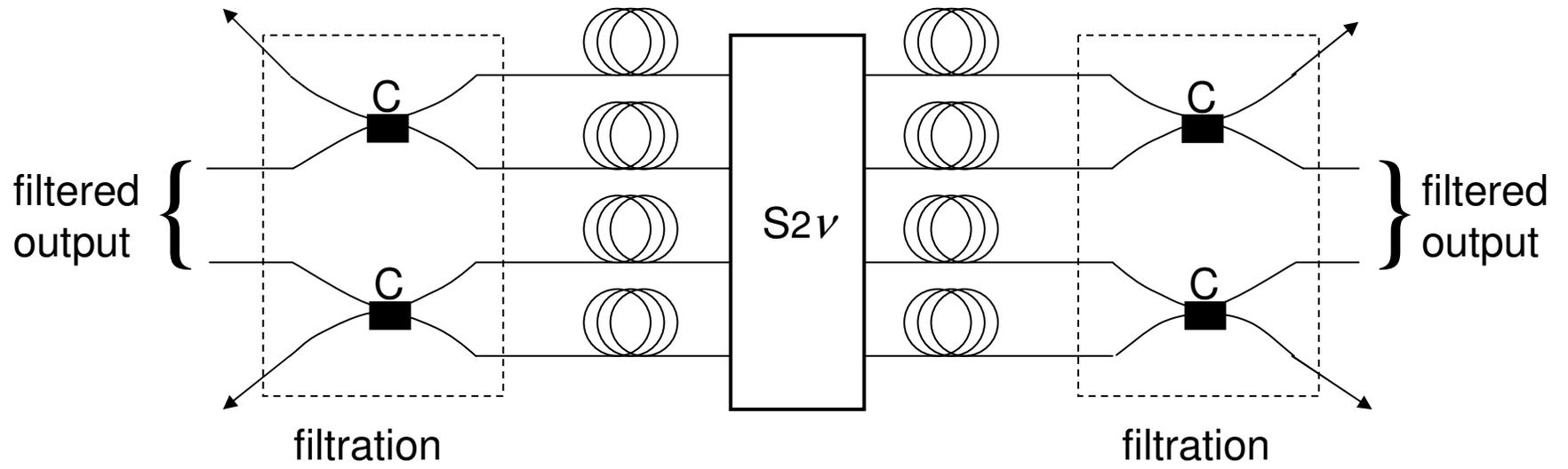
      {\bf Fig. 3.} 
      Implementation of error filtration for entangled particles using 
      multiple optical fibers. 
      A source ($S_{2\nu}$) produces two photons in the entangled state 
      $(|1\rangle_A |1\rangle_B + |2\rangle_A |2\rangle_B + |3\rangle_A 
      |3\rangle_B + |4\rangle_A |4\rangle_B )/2$. States $|i\rangle_{A(B)}$ 
      travel to receivers A(B) through different optical fibers. As in 
      Fig. 1 the transmission states will be affected by independent phase 
      noise. The receivers filter out the noise by projecting the state onto 
      the subspaces spanned by $|1\rangle_{A(B)}+|2\rangle_{A(B)}$, 
      $|3\rangle_{A(B)}+|4\rangle_{A(B)}$ using 50/50 couplers. When the 
      projections of both parties succeed the noise has been filtered 
      out. When the projection of either of the parties fails the state is 
      rejected. 
      Note that this decoding operation is based on the Hadamard 
      transform and is distinct from the decoding operation based on the 
      Fourier transform considered in the main text. However one can easily 
      show, see the supplementary material, that both methods filter out the 
      same amount of noise. 

      \bigskip 
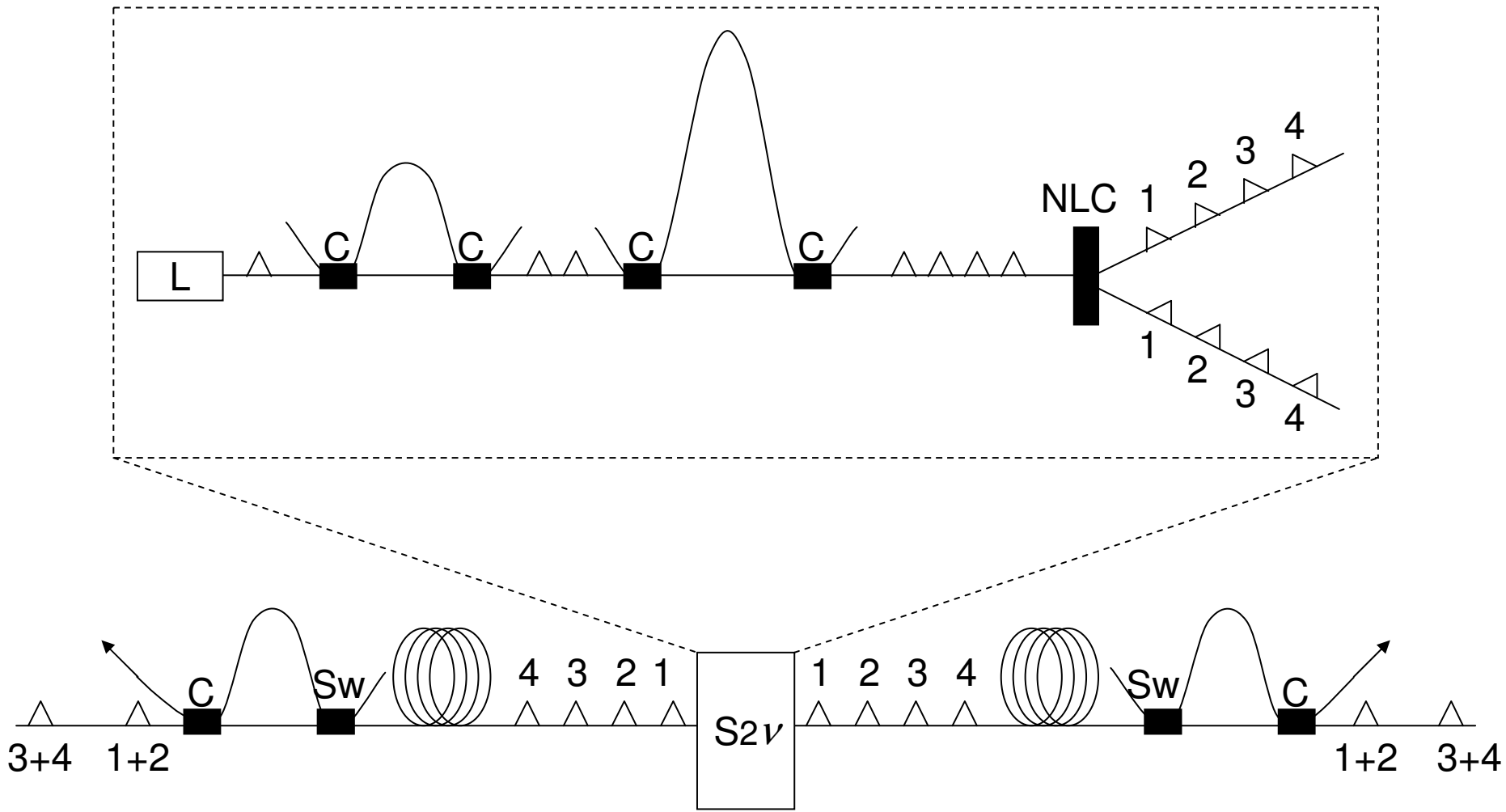
      \noindent {\bf Fig. 4} Implementation of error filtration for 
      entangled particles using time bins. A source ($S_{2\nu}$) 
      produces two photons in the entangled state $(|1\rangle_A 
      |1\rangle_B + |2\rangle_A |2\rangle_B + |3\rangle_A |3\rangle_B + 
      |4\rangle_A |4\rangle_B )/2$ where states $|i\rangle_{A(B)}$ 
      correspond to different time bins traveling through the same 
      optical fiber. A possible such source, adapted from 
      \cite{2photonsource99}, is described in the inset: a laser (L) 
      produces intense pulses of light which pass through two unbalanced 
      Mach-Zender (MZ) interferometers so as to produce 4 coherent 
      equally spaced pulses. The pulses impinge on a non linear crystal 
      (NLC) thereby producing the entangled state by parametric down 
      conversion. As in Fig. 2 the transmission states will be affected 
      by independent phase noise. To filter out the noise, the photons 
      are sent through MZ interferometers. A switch (Sw), synchronized 
      with the source, sends time bins 2 and 4 through the long arm and 
      time bins 1 and 3 through the short arm. Time bins 1 and 2 and 
      time bins 3 and 4 then interfere. The state is kept if both 
      photons exit through the lower branch, in which case it has been 
      projected onto the subspace spanned by 
      $|1\rangle_{A(B)}+|2\rangle_{A(B)}$, 
      $|3\rangle_{A(B)}+|4\rangle_{A(B)}$. If either of the photons 
      exit through the upper branch the filtration has failed. As 
      in Fig. 3 this example is based on the Hadamard transform.

      \end{document}